\documentclass[a4paper,fleqn,usenatbib,useAMS]{mnras}

\usepackage[T1]{fontenc}
\usepackage{ae,aecompl}

\usepackage{graphicx}	
\usepackage{amsmath}	
\usepackage{amssymb}	
\usepackage{hyperref}
\usepackage{ae,aecompl}
\usepackage{float}
\usepackage{longtable}
\usepackage{cite}
\usepackage{hyperref}
\usepackage{natbib}
\usepackage{nccmath}
\usepackage{multirow}
\usepackage{threeparttable}

\usepackage{float}
\usepackage{caption}
\usepackage{subcaption}
\usepackage{mwe}
\usepackage{enumitem}

\title[A diagnostic tool for SNRs identification]{A diagnostic tool for the identification of Supernova Remnants}

\author[M. Kopsacheili et al.]{
M. Kopsacheili,$^{1,2}$\thanks{E-mail: mariakop@physics.uoc.gr}
A. Zezas,$^{1,2,3}$
I. Leonidaki$^{1,2}$
\\
$^{1}$ Physics Department and Institute of Theoretical and Computational Physics, University of Crete, 71003 Heraklion, Crete, Greece\\
$^{2}$ Foundation for Research and Technology-Hellas, 71110 Heraklion, Crete, Greece\\
$^{3}$ Harvard-Smithsonian Center for Astrophysics, 60 Garden Street, Cambridge, MA 02138, USA
}

\date{Accepted XXX. Received YYY; in original form ZZZ}

\pubyear{2015}

\begin{document}
\label{firstpage}
\pagerange{\pageref{firstpage}--\pageref{lastpage}}
\maketitle

\begin{abstract}
We present new diagnostic tools for distinguishing supernova  remnants (SNRs) from HII regions. Up to now, sources with flux ratio [S II]/H$\rm{\alpha}$ higher than 0.4 have been considered as SNRs. Here, we present the combinations of three or two line ratios as  more effective tools for the separation of these two kinds of nebulae, depicting them as 3D surfaces or 2D lines. The diagnostics are based on photoionization and shock excitation models (MAPPINGS III) analysed with Support Vector Machine (SVM) models for classification. The line-ratio combination that gives the most efficient diagnostic is: [O I]/H$\rm{\alpha}$ - [O II]/H$\rm{\beta}$ - [O III]/H$\rm{\beta}$. This method gives $98.95\% $ completeness in the SNR selection and $1.20\%$ contamination. We also define the [O I]/H$\rm{\alpha}$ SNR selection criterion and we measure its efficiency in comparison to other selection criteria.

\end{abstract}



\section{Introduction}

Study of SNR demographics and their physical properties (density, temperature, shock velocities) is very important in order to understand their role in galaxies. Their feedback to the Interstellar Medium (ISM), and consequently to the entire galaxy, is of high importance since they provide significant amounts of energy that heat the ISM and they enrich it with heavy elements. They are fundamentally related to the star-forming process in a galaxy, inasmuch as the compression of the ISM by the shock wave, under appropriate conditions, can lead to the formation of new stars. Having a complete census of SNR populations can also give us a picture of the on-going massive star formation rate (SFR) since they depict the end points of massive stars ($M > 8M\odot $).

\par Many photometric and spectroscopic studies of SNRs, have been carried out in our Galaxy (e.g. \citealt{Dan2013}; \citealt{Boumis2009}) but also in extragalactic environments (e.g. \citealt{Vucetic2015}; \citealt{Leonidaki2013}; \citealt{Leonidaki2010}; \citealt{Blair1997}). These studies increase the number of known SNRs, and also provide significant information about their physical and kinematic properties, as well as information on their interaction with their local ISM. According to \citet{Green2017} the known number of the optical Galactic SNRs is 295, while most of the studies on extragalactic environments present a few dozen SNRs per galaxy, except for a handfull of extreme cases: e.g. M83 with 225 photometric SNRs;  (\citealt{BlairWinklerLong2013}; \citealt{BlairWinklerLong2012}), M33 with 220 (\citealt{Long2018}), M31 with 150 (\citealt{Lee2014}). The small number of observed extragalactic SNRs compared to those in our Galaxy, is the result of different sensitivity limits and also different selection criteria. The identification of SNRs in our Galaxy or the Magellanic clouds is generally based on the detection of extended non-thermal radio sources, or X-ray sources, while studies in other galaxies rely on photometric or spectroscopic measurements of diagnostic spectral lines.  
\par 
The most common means of identifying SNRs in the optical regime, is the use of the flux ratio of the [S II] ($\lambda\lambda 6717, 6731$) to H$\rm{\alpha}$ ($\lambda 6563$) emission lines, as first suggested by \citet{Mathewson&Clarke} based on studies of SNR population in the Large Magellanic Cloud (LMC). Usually, nebulae with [S II]/H$\rm{\alpha}$ ratio higher than 0.4 are considered as SNRs. Indeed, we expect SNRs to give higher values of [S II] than HII regions since collisionally excited $S^+$ behind the shock front gives strong [S II] emission, while in HII regions sulphur is mostly in the form of $S^{++}$. However, within the years, this low limit for the [S II]/H$\rm{\alpha}$ ratio has been slightly modified in order to take into account different interstellar densities for the [S II]/H$\rm{\alpha}$ ratio (\citealt{Daltabuit_DOdorico_Sabbadin}), different galaxy metallicities (\citealt{Leonidaki2013}; \citealt{DOdorico_Benvenuti_Sabbadin}), difficulties in distinguishing SNRs from HII regions on the borderline between them (\citealt{Fesen1985}) or strong emission from [N II] (\citealt{Dopita2010}).  Consequently, a more robust diagnostic tool seems to be necessary. \citealt{Fesen1985} recognizing this need, suggested the line ratios of [O I]/H$\rm{\beta}$ and [O II]/H$\rm{\beta}$ that seem to efficiently differentiate SNRs from HII regions. 
\par Advanced observing techniques (multi-slit spectroscopy) give us the ability to obtain full spectral information for large numbers of sources. This, in combination with the development of advanced photoionization and shock models (\citealt{kewly2000}; \citealt{Allen2008}) allows us to examine more accurately spectral features of nebulae and compare data with theory. Several studies have used diagnostic diagrams to separate objects based on their excitation mechanisms, like HII regions and active galactic nuclei (AGN) using 2D or multi-D diagnostics (e.g. \citealt{Stampoulis2019}; \citealt{Souza2017}; \citealt{Vogt2014}; \citealt{kewly2000}).
\par  Our study focuses on diagnostic diagrams that separate SNRs from HII regions.  We present a set of new diagnostic tools for the identification of optical SNRs. 
These models allow us to derive theory-driven diagnostics that overcome the limitation of the empirical diagnostics employed so far.

\par  The outline of this paper is as follows. In Section 2, we describe the models we used, in Section 3, we talk about the emission line ratios that we examined, the classification method and the most accurate line ratio combination and in Section 4, we discuss our results.

\section{Models}
In order to generate an emission-line diagnostic tool that is able to separate SNRs from HII regions, we used the results from MAPPINGS III, a photoionization code (\citealt{Groves2004}; \citealt{Dopita_Groves}; \citealt{Sutherland_Dopita}; \citealt{Binette1985}) that predicts emission-line spectra of a medium that is subject to photoionization or shock excitation (\citealt{Allen2008}). We obtained the line ratios from the compilation of photoionization and shock excitation model grids available in the ITERA (IDL Tool for Emission-line Ratio Analysis) tool (\citealt{ITERA1}; \citealt{ITERA2}).
 \\
\\
\textbf{Starburst models}\\
ITERA includes two sets of starburst models, i.e. emission-line spectra emerging from gas photo-ionized by two different sets of stellar population models. These correspond to the spectra expected from HII regions or star-forming galaxies.
\begin{enumerate}[label=(\roman*)]
\item{Kewley2000: The first set of models are from \citet{kewly2000}. These are photoionizaton models based on stellar ionizing spectra created either by the PEGASE-2 (\citealt{PEGASE2}) or the Starburst99 (\citealt{Starburst99}) stellar population synthesis codes and under two star-formation scenarios: continuous and instantaneous. The ITERA library contains MAPPINGS III models for various values of the ionization parameter (ranging from $2\times 10^5$ to $4\times 10^ 8 \rm{cm\, s^{-1}}$) and metallicities (from $0.01$ to $3\, \rm{Z_\odot} $ for PEGASE-2, and from $0.05$ to $2\, \rm{Z_\odot} $ for Starburst99).}

\item {Levesque09: The second set of models is from \citet{levesque2010}. These are stellar photoionization models with an updated version of Starburst99 code (\citealt{vazquez_leitherer2005}) with continuous star formation and instantaneous burst models, extending to a wider range of ages (0-10 Myr) and  examining not only the case of standard but also of high mass-loss tracks, which better approximate the mass loss of massive stars (\citealt{levesque2010})}.
\end{enumerate} 

\noindent
\textbf{Shock models}\\
\citet{Allen2008} provide a library of spectral line intensities for shock models of different velocities, magnetic parameters, abundances and densities, with and without a photoionizing precursor. From these models we used these that combine the emission from the pre and post-shocked regions which better represent the observation of unresolved sources. They cover velocity ranges from 100-1000 $\rm{km\,s^{-1}}$ and magnetic parameters ($\rm{B/n^{1/2}}$, where B is the transverse components of the preshock magnetic field and n is the preshock particle number density) from $\rm{10^{-4}}$ to $\rm{10\mu\, G\, cm^{3/2}}$. They also consider different abundances (LMC, SMC, solar, twice solar and \citealt{Dopita2005} solar abundance) and densities ($\rm{n = 0.01, 0.1, 1.0, 10, 100, 1000\, cm^{-3} }$).\\
\\
From these models, we have two sets of line ratios involving the most prominent optical lines for the different abundance and ISM conditions considered. One set applies to SNRs (resulting from the shock models) and another one refers to HII regions (resulting from the starburst models).  In the case of shock models, each point of the set is characterized by a shock velocity, a magnetic parameter, a density and an abundance, while for starburst models it is described by the ionization parameter, age, abundance and density.

\section{Optimal combination of lines}
In order to find the optimal emission line ratios that best distinguish SNRs from HII regions, we examined two and three-dimensional diagnostics involving different lines for the full range of abundances and densities. The emission lines we opted to use are various forbidden lines, which tend to be stronger in shock-excited than in photoionized regions: [N II]($\lambda 6583$), [S II]($\lambda\lambda 6716,6731$), [O I]($\lambda 6300$), [O II]($\lambda\lambda 3727, 3729$), [O III]($\lambda 5008$).\par
\begin{figure}
\centering
 \includegraphics[width=0.5\textwidth]{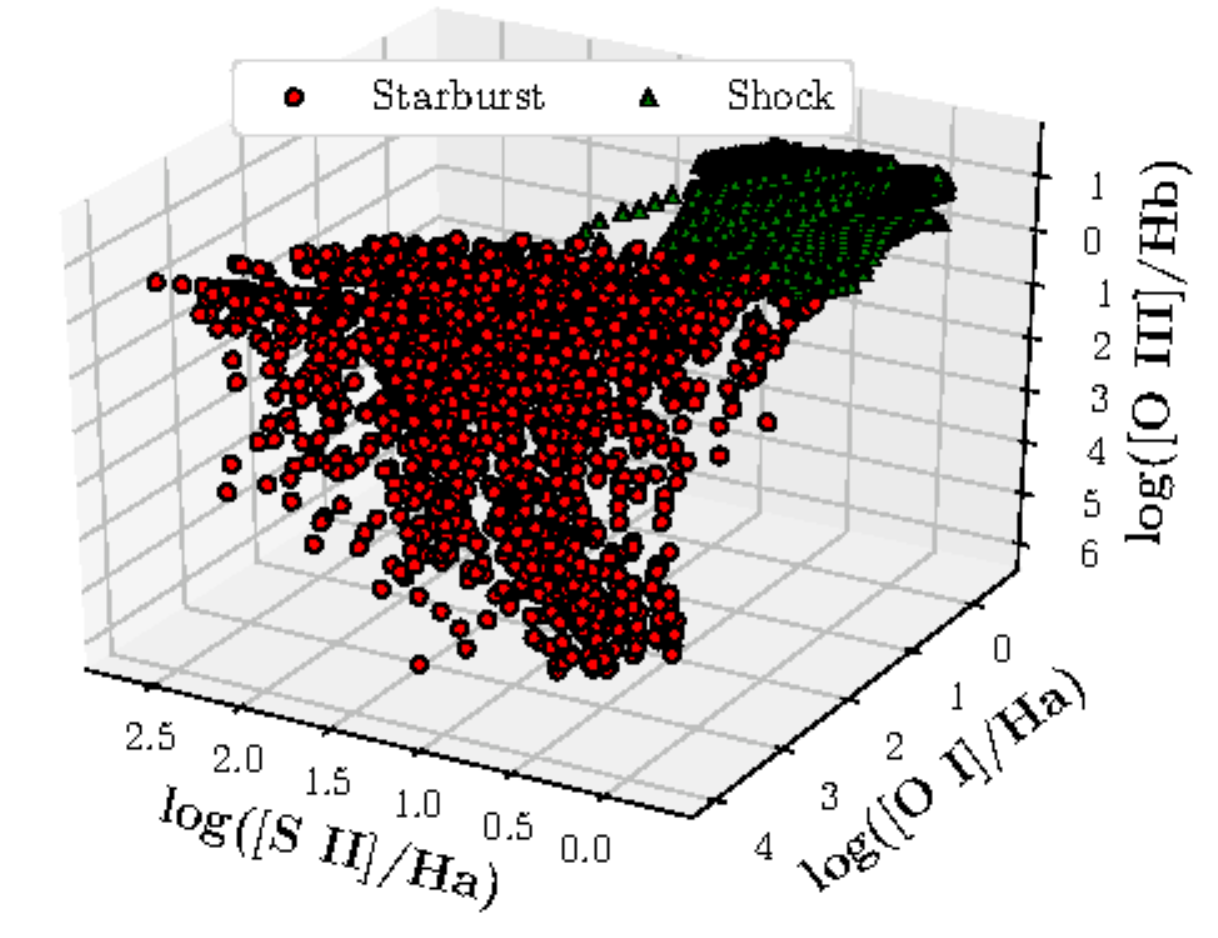}
\caption{\label{fig:SII_OI_OIII_only} A 3D diagram for the [S II]/H$\rm{\alpha}$ - [O I]/H$\rm{\alpha}$ - [O III]/H$\rm{\beta}$ line ratios. The points show the expected line ratios for starburst models (red) and shock models (green) for different ISM parameters based on the ITERA library of MAPPINGS III model.}
\end{figure}

 For example \autoref{fig:SII_OI_OIII_only} presents a 3D diagram of the line ratios  [S II]/H$\rm{\alpha}$ - [O I]/H$\rm{\alpha}$ - [O III]/H$\rm{\beta}$ for shock models (representing SNRs; green triangles) and starburst models (representing HII regions; red circles). As we can see shock models are quite well separated from starburst models, although there is some overlap.

\subsection{Definition of the diagnostic}

Aiming to quantify this separation, we construct a 2D-curve (using 2 line ratios) or a 3D-surface (using 3 line ratios) that optimally distinguishes SNRs from HII regions. In order to find the most appropriate separating surface, we used the support vector machine (SVM) models. Specifically we used the python module scikit-learn \footnote{https://scikit-learn.org/stable/modules/svm.html}, a set of supervised learning algorithms for classification, that separates a set of data in two or more classes. SVM can classify different classes of objects on the basis of separating surfaces  in the multi-dimensional space defined by characteristic parameters of these objects (here the line ratios). This boundary can be described by a function of two or more variables, depending on the dimensionality of the input data (here, the number of line ratios which are used). The function of this boundary (decision function) has the following form (e.g. \citealt{stat_book}):
\begin{ceqn}
\begin{align*}
   \sum_{i=1}^{n} {\alpha _i y_iK(x^T,x_i)} + \rho
\end{align*}
\end{ceqn}
where $n$ is the number of the support vectors, (i.e. the points nearest to the distance of the closest point from either class), y is the class, $\alpha$ is the lagrangian multiplier vector, $\rho$ is the intercept term and $K(x^T,x_i)$ is the kernel function. The general form of the kernel function is $K(x^T,x_i)=(\gamma<x,x_i> +r)^d$, where $\gamma$ is the kernel width parameter, $x_i$ are the support vectors, $r$ is a constant coefficient, which in our case equals to 1, and d is the degree of the polynomial.  We explored various values of $\gamma$, from 0.2-1.0, and we selected the ones that better discriminate between different classes, as we explain next. We examined two cases of kernel functions, linear (d=1) and polynomial (d=3), the latter giving more flexibility in the case of complex separating lines/surfaces, for all combinations of two or three of the line ratios [N II]/H$\rm{\alpha}$, [S II]/H$\rm{\alpha}$, [O I]/H$\rm{\alpha}$, [O II]/H$\rm{\beta}$, [O III]/$\rm{\beta}$.

\par
\begin{figure*}
\begin{minipage}{160mm}
\centering
 \vspace*{5pt}%
 \hspace*{\fill}%
  \begin{subfigure}{1.0\textwidth}     
    \centering
    \includegraphics[width=\textwidth]{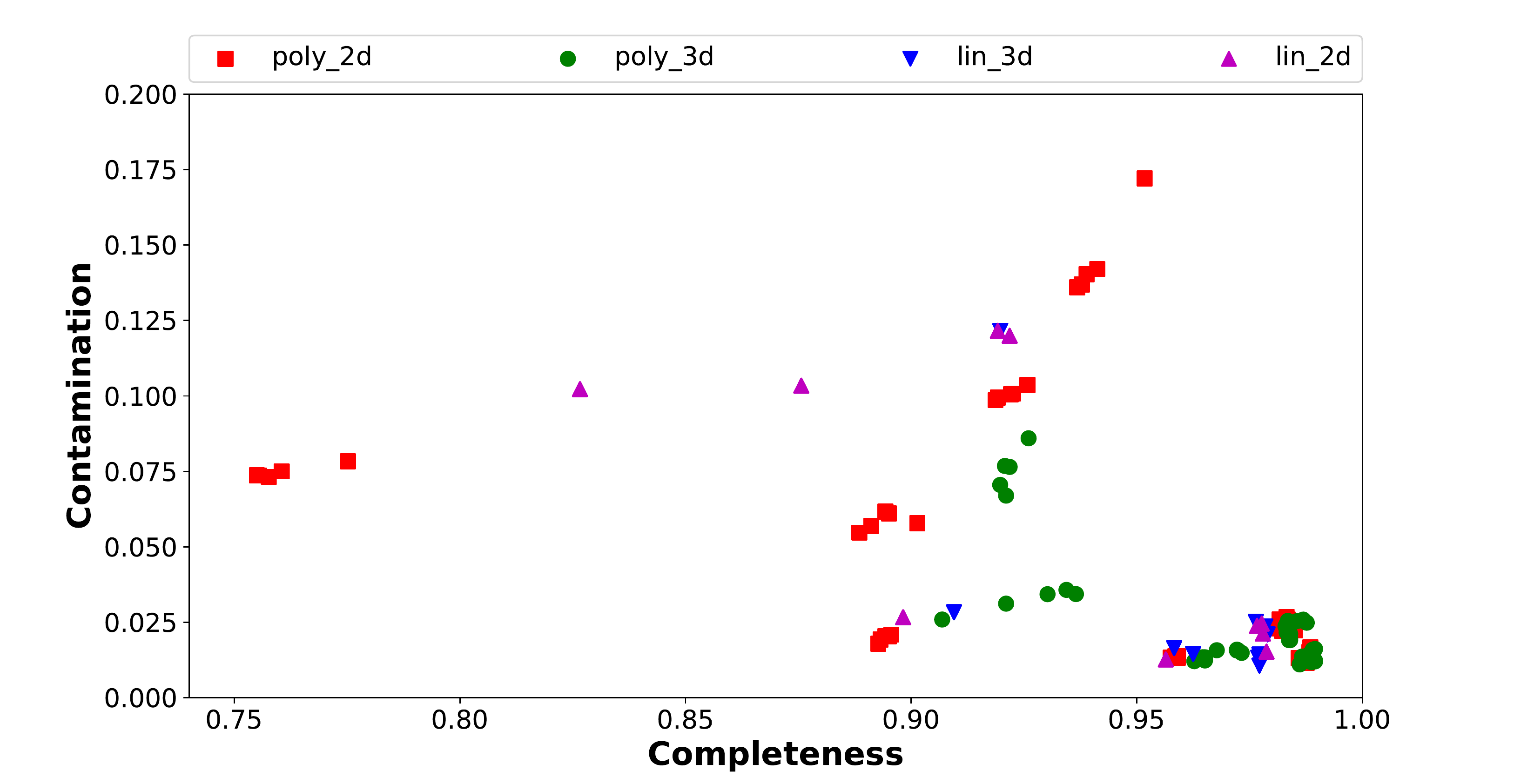}%
  \end{subfigure}

  \vspace*{5pt}%

  \hspace*{\fill}%
   \begin{subfigure}{1.0\textwidth}        
    \centering
    \includegraphics[width=\textwidth]{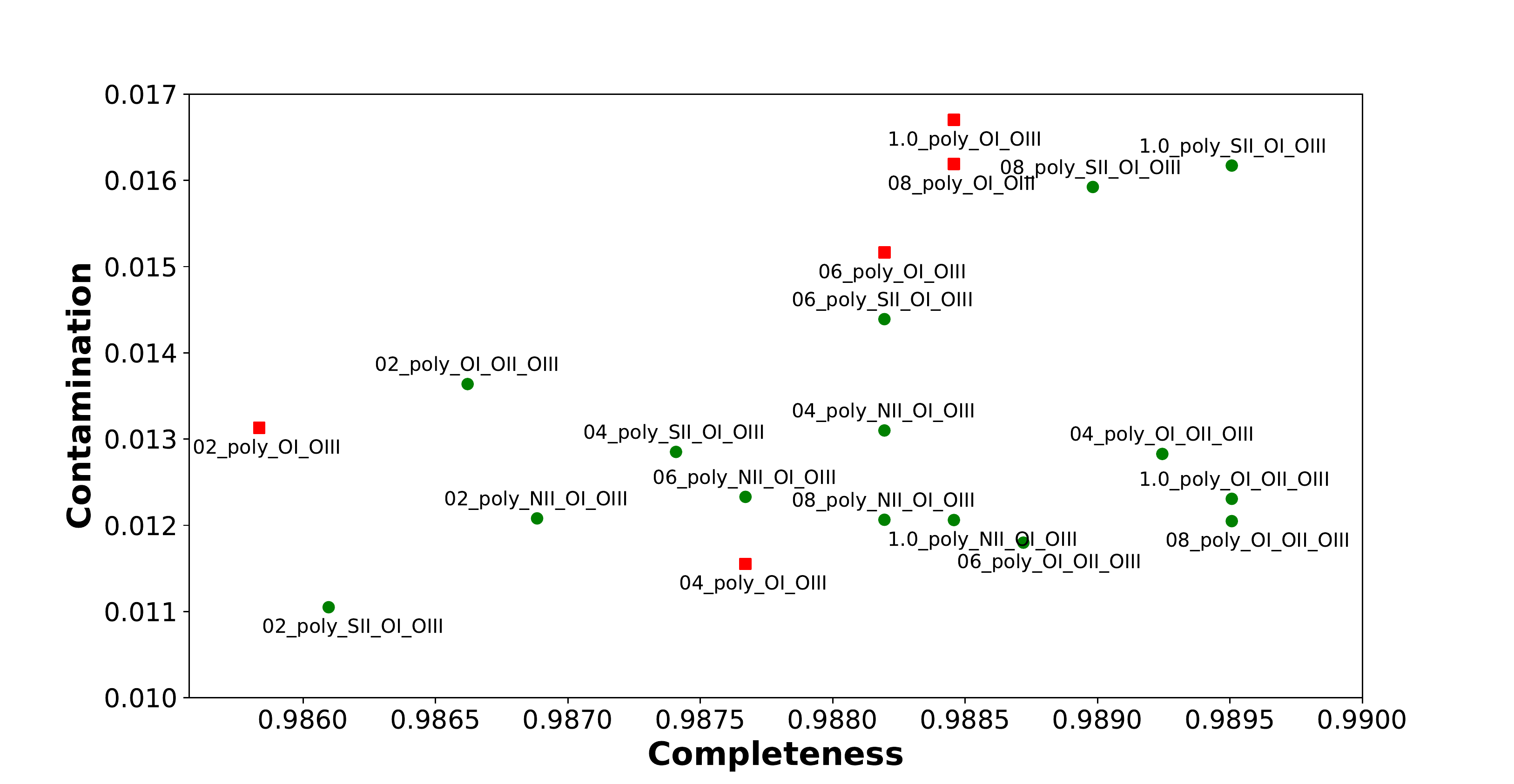}%
 \end{subfigure}
\caption{\label{fig:cp_ct_all} \small{Completeness versus contamination for each diagnostic. Different colors and shapes correspond to different kernel of the decision function: the red squares refer to 2D diagnostics with polynomial kernel in the decision function (poly{\_}2d), the green circles to 3D diagnostics with polynomial kernel (poly{\_}3d), the blue triangles-down to 3D diagnostics with linear kernel (lin{\_}3d) and the purple triangles-up to 2D diagnostics with linear kernel (lin{\_}2d). The bottom panel shows more clearly the high completeness - low contamination region of the top panel. Each point is labeled as $\gamma$\_kernel\_lines, where $\gamma$ is the kernel width parameter (e.g $\gamma$ = 0.2 corresponds to 02 kernel), poly refers to polynomial kernel of the decision function and then follow the emission lines ratio used in each diagnostic indicated by the forbidden line involved, e.g. with SII we refer to [S II]/H$\rm{\alpha}$ etc.}}
\end{minipage}
\end{figure*}

{Since we are interested in the definition of a diagnostic tool for SNRs we consider: (a) the completeness of shock models (i.e. SNRs) defined as the number of true positives over the sum of true positives and false negatives (i.e. the total number of shock models) and (b) their contamination by starburst models (i.e. HII regions) that is the number of false positives over the sum of true and false positives. The line ratios that maximize the completeness and minimize the contamination are those that we consider as the best diagnostics for distinguishing SNRs from HII regions. In \autoref{fig:cp_ct_all} (top) we see the completeness versus the contamination for each line combination and kernel of different functional form (linear or polynomial). The line combinations with the highest completeness and lowest contamination (i.e. the bottom right region of \autoref{fig:cp_ct_all}) are shown in the bottom panel of \autoref{fig:cp_ct_all}.}

\subsection{Optimal diagnostics}
From these results it is clear that the polynomial kernel is  more efficient than the linear kernel and hence all the diagnostics we present use a polynomial kernel in their decision function. As we see in \autoref{fig:cp_ct_all} (bottom panel), the most effective line combination is the ([O I]/H$\rm{\alpha}$ - [O II]/H$\rm{\beta}$ - [O III]/H$\rm{\beta}$) with $\gamma$ = 0.8 (diagnostic A). At the same time, the combination ([S II]/H$\rm{\alpha}$ - [O I]/H$\rm{\alpha}$ - [O III]/H$\rm{\beta}$) with $\gamma$ = 1.0 (diagnostic B) works very well and it can be used in cases where the wavelength range of the spectra (or narrow band imaging) is limited to redder wavelengths. We obtain similar results with the line ratios ([N II]/H$\rm{\alpha}$ - [O I]/H$\rm{\alpha}$ - [O III]/H$\rm{\beta}$) with $\gamma$ = 1.0 (diagnostic C). 
\par In two dimensions the most effective diagnostic is the combination of line ratios ([O I]/H$\rm{\alpha}$ - [O III]/H$\rm{\beta}$) with $\gamma$ = 0.4 (diagnostic D). If we are restricted in the red and the blue parts of the spectrum, the most powerful diagnostics are ([N II]/H$\rm{\alpha}$ - [O I]/H$\rm{\alpha}$) with $\gamma$ = 1.0 (diagnostic E) and ([O II]/H$\rm{\beta}$ - [O III]/H$\rm{\beta}$) with $\gamma$ = 0.2 (diagnostic F) respectively. 
\par \autoref{table:CP_CT_table} shows the completeness (CP) and the contamination (CT) for each one of these diagnostics. CPs and CTs show that in general 3D  give more accurate results that 2D diagnostics.
The rest of the diagnostics we examined are presented in the Appendix.
\begin{table}
\caption{\small{Completeness and contamination for the diagnostics described in \S 3.2.}}
\begin{threeparttable}
\begin{tabular}{|p{1.2cm}|p{0.7cm}|p{0.7cm}|p{0.7cm}|p{0.7cm}|p{0.7cm}|p{0.7cm}|}
\hline
Diagnostics:  & \hfil  A & \hfil  B & \hfil  C & \hfil  D & \hfil  E & \hfil  F \\
\hline
 Compl.   & 0.990 & 0.990 & 0.989 & 0.988 & 0.983 & 0.901    \\
 Cont.    & 0.012 & 0.016 & 0.012 & 0.012 & 0.023 & 0.058    \\
\hline
\end{tabular}
\begin{tablenotes}
 \item \scriptsize A: [O I]/H$\rm{\alpha}$ - [O II]/H$\rm{\beta}$ - [O III]/H$\rm{\beta}$\\
B: [S II]/H$\rm{\alpha}$ - [O I]/H$\rm{\alpha}$ - [O III]/H$\rm{\beta}$\\
C: [N II]/H$\rm{\alpha}$ - [O I]/H$\rm{\alpha}$ - [O III]/H$\rm{\beta}$\\
D: [O I]/H$\rm{\alpha}$ - [O III]/H$\rm{\beta}$\\
E: [N II]/H$\rm{\alpha}$ - [O I]/H$\rm{\alpha}$\\
F: [O II]/H$\rm{\beta}$ - [O III]/H$\rm{\beta}$
\end{tablenotes}
\end{threeparttable}
\label{table:CP_CT_table}
\end{table}
\par
\begin{figure}
\centering
 \vspace*{5pt}%
 \hspace*{\fill}%
  \begin{subfigure}{0.5\textwidth}     
    \centering
    \includegraphics[width=\textwidth]{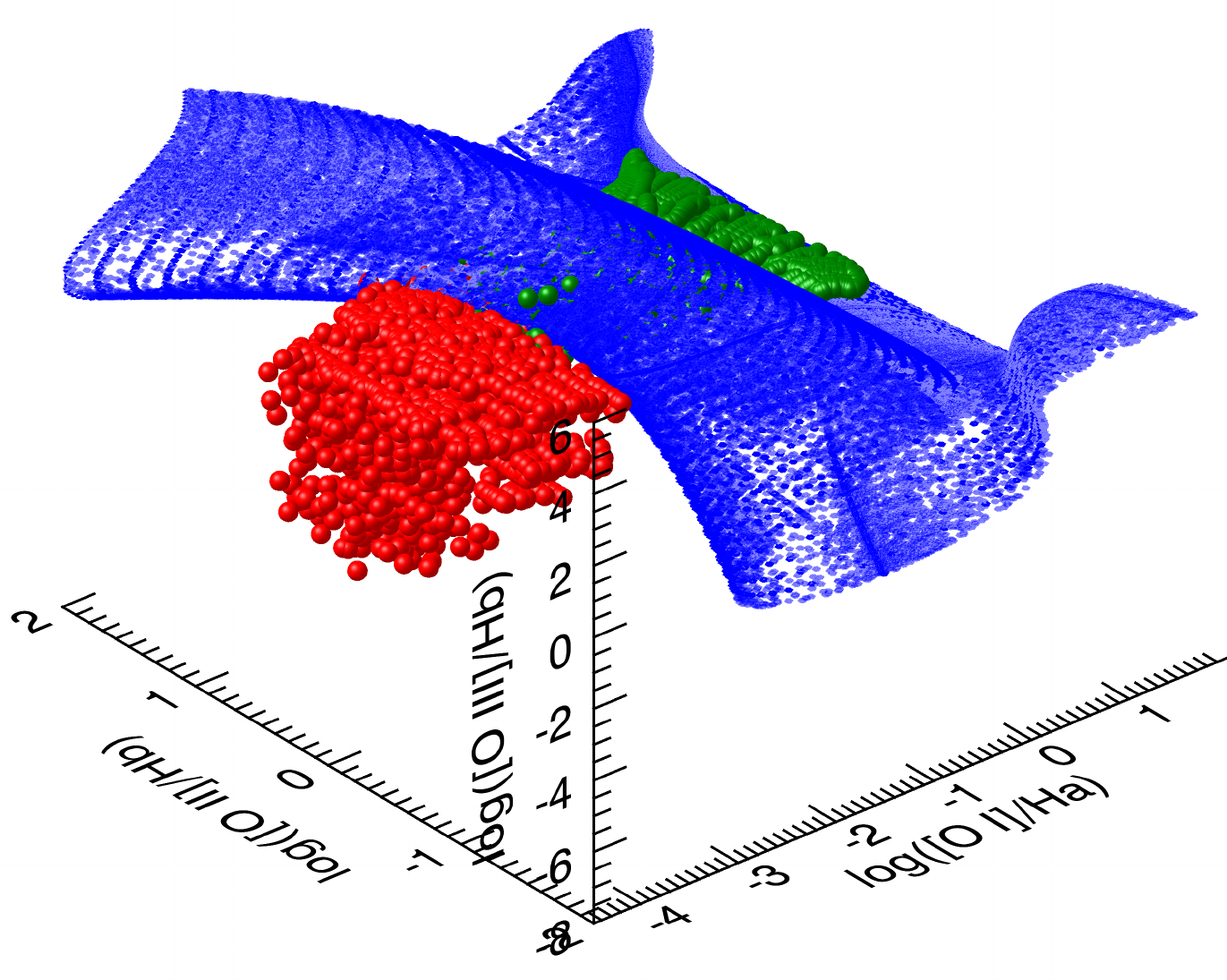}%
  \end{subfigure}

  \vspace*{5pt}%

  \hspace*{\fill}%
   \begin{subfigure}{0.5\textwidth}        
    \centering
    \includegraphics[width=\textwidth]{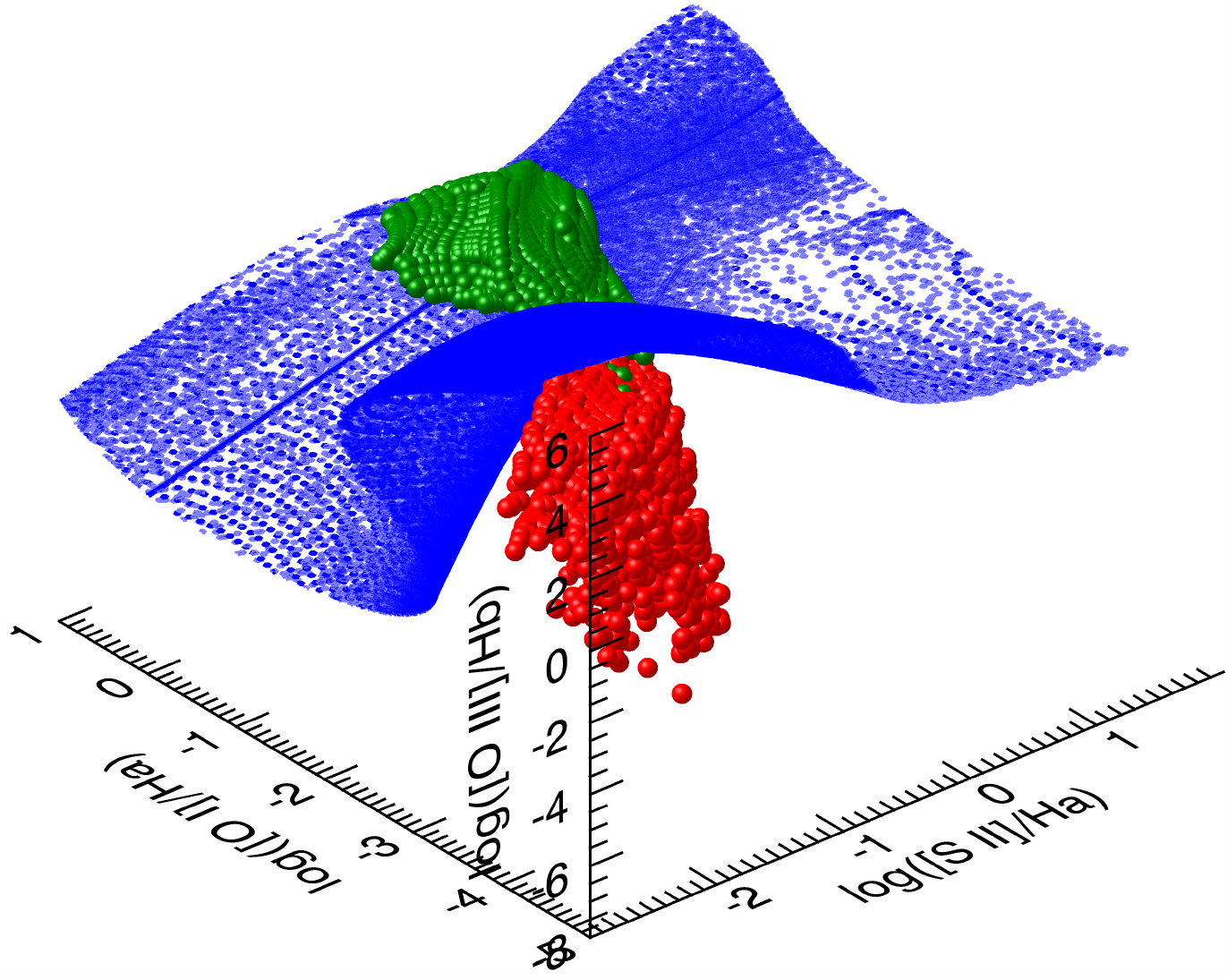}%
 \end{subfigure}
 
   \vspace*{5pt}%

  \hspace*{\fill}%
   \begin{subfigure}{0.5\textwidth}        
    \centering
    \includegraphics[width=\textwidth]{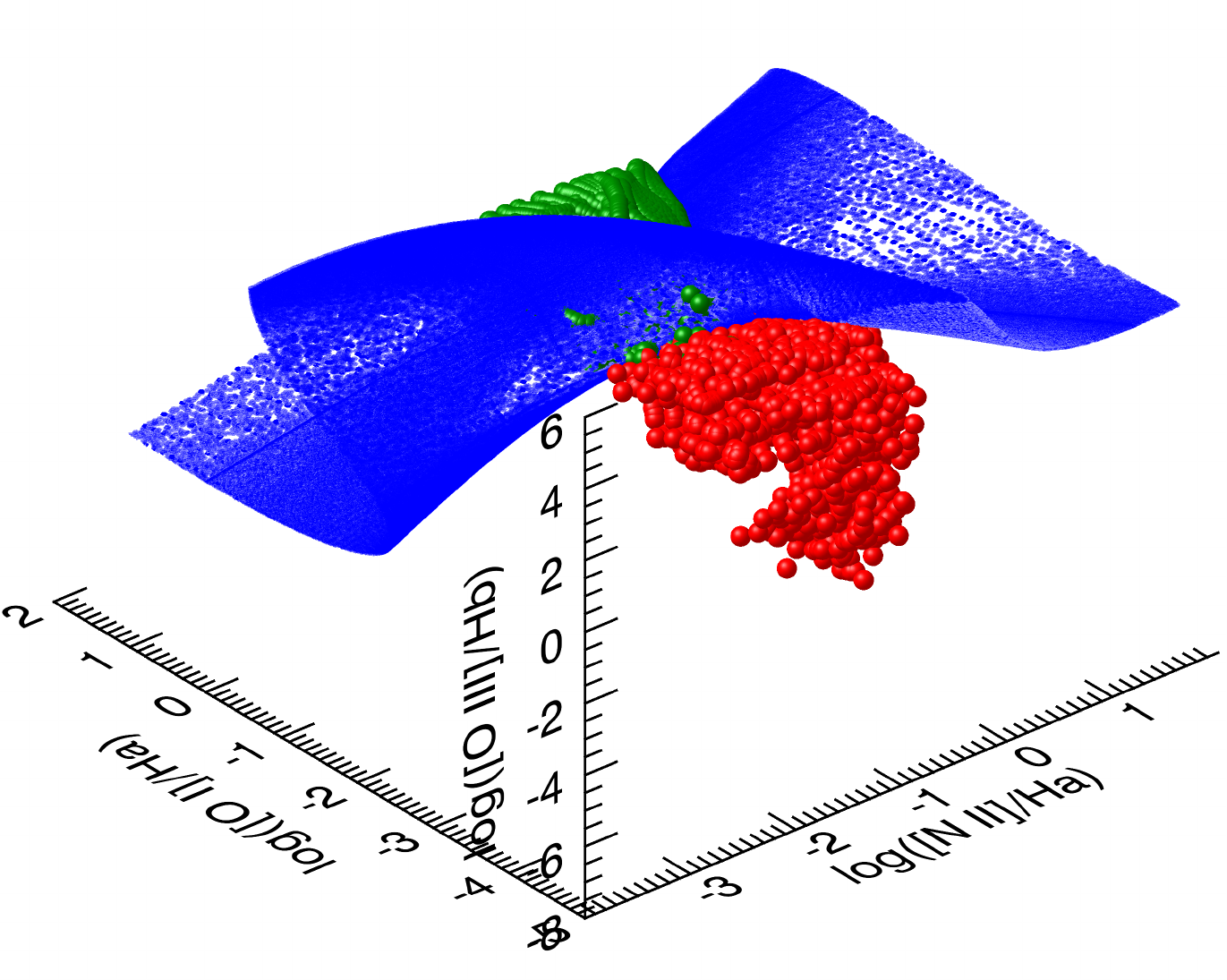}%
 \end{subfigure}%
\caption{{\label{fig:ed_surf} \small{The surfaces separating shock models (SNRs, green) from starburst models (HII regions, red) for the diagnostics A ([O I]/H$\rm{\alpha}$ - [O II]/H$\rm{\beta}$ - [O III]/H$\rm{\beta}$), B ([S II]/H$\rm{\alpha}$ - [O I]/H$\rm{\alpha}$ - [O III]/H$\rm{\beta}$) and C ([N II]/H$\rm{\alpha}$ - [O I]/H$\rm{\alpha}$ - [O III]/H$\rm{\beta}$), from top to bottom respectively. The data points are drawn from the ITERA compilation of MAPPINGS III models for representative densities and metallicities (\S 2). Animations showing the rotation of these surfaces are available in the on-line version.}}}
\end{figure}

\begin{figure}
\centering
 \vspace*{5pt}%
 \hspace*{\fill}%
  \begin{subfigure}{0.5\textwidth}     
    \centering
    \includegraphics[width=\textwidth]{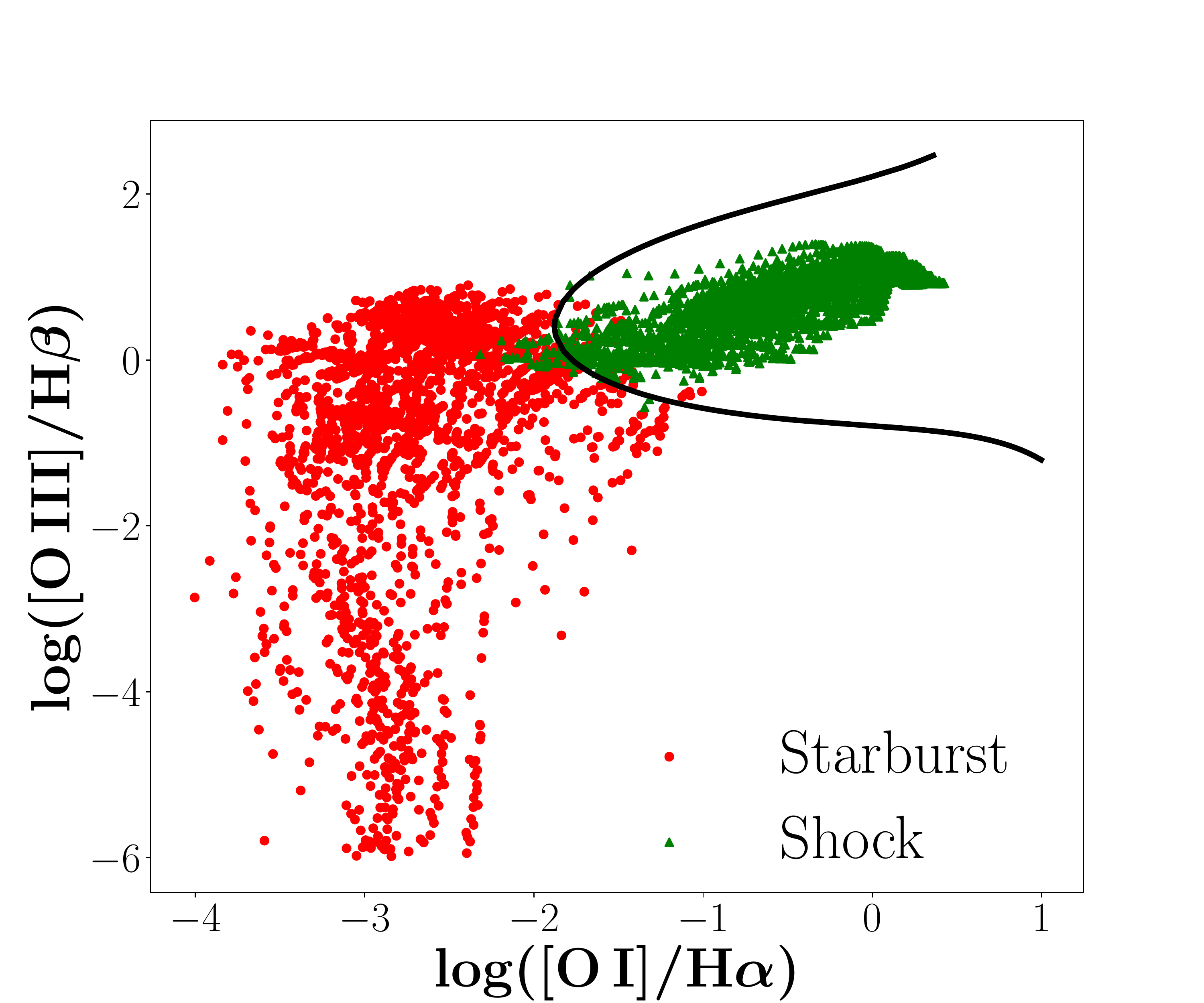}%
  \end{subfigure}

  \vspace*{5pt}%

  \hspace*{\fill}%
   \begin{subfigure}{0.5\textwidth}        
    \centering
    \includegraphics[width=\textwidth]{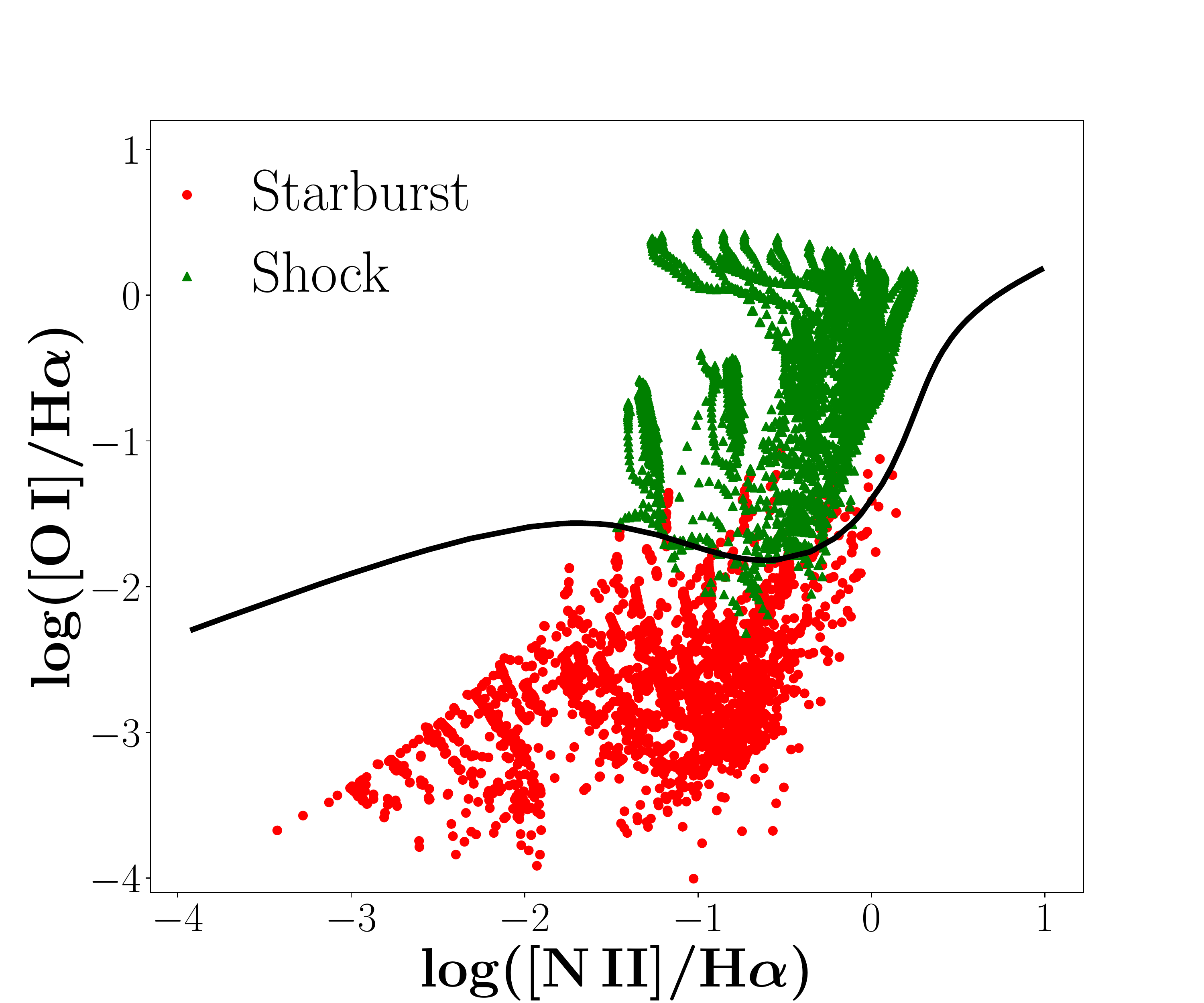}%
 \end{subfigure}
 
   \vspace*{5pt}%

  \hspace*{\fill}%
   \begin{subfigure}{0.5\textwidth}        
    \centering
    \includegraphics[width=\textwidth]{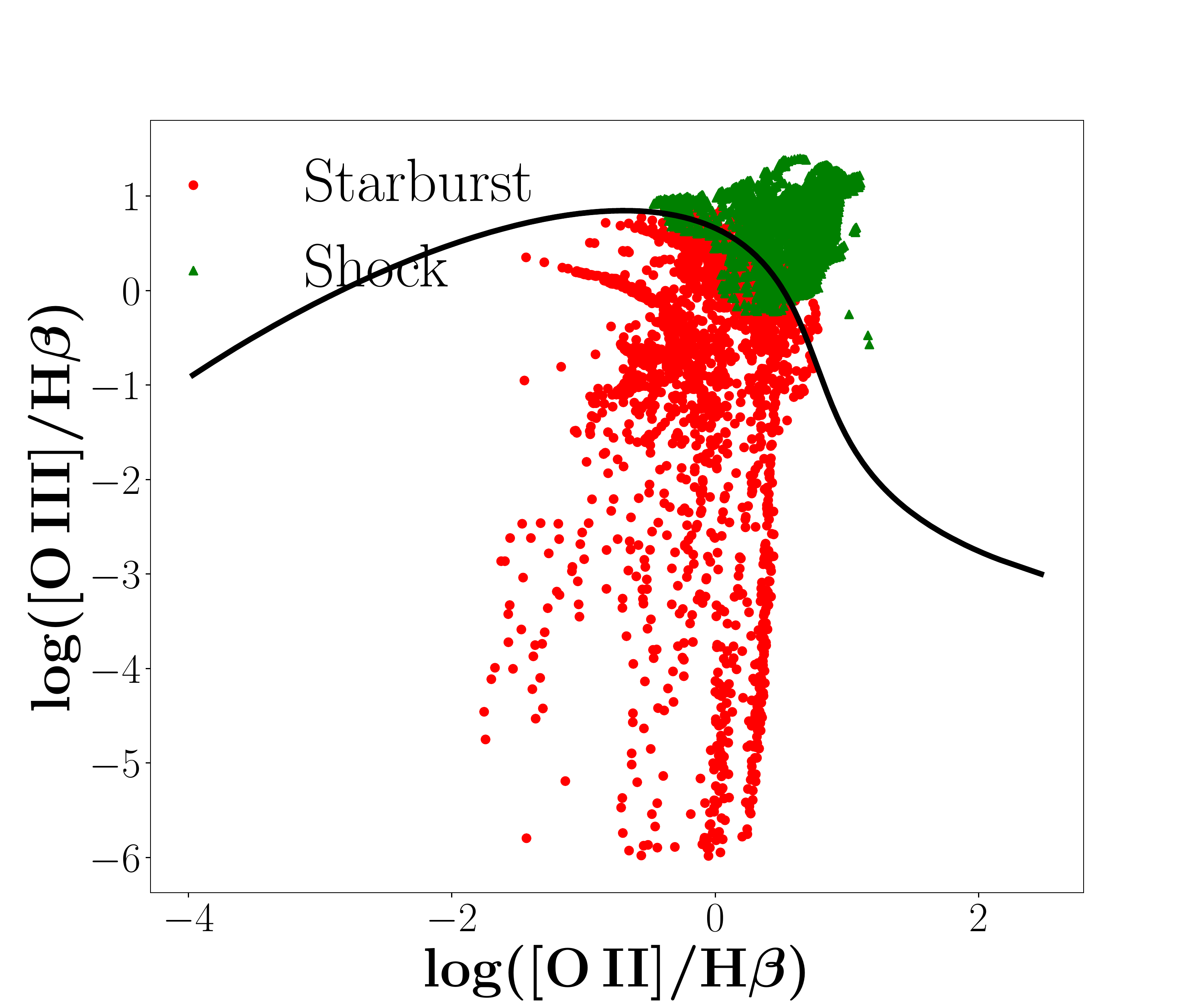}%
 \end{subfigure}%
\caption{{\label{fig:OI_OIII} \small{The lines that separate shock models (SNRs, green) from starburst models (HII regions, red) for the diagnostics D ([O I]/H$\rm{\alpha}$ - [O III]/H$\rm{\beta}$), E ([N II]/H$\rm{\alpha}$ - [O I]/H$\rm{\alpha}$) and F ([O II]/H$\rm{\beta}$ - [O III]/H$\rm{\beta}$), from top to bottom respectively. The data points are drawn from the ITERA compilation of MAPPINGS III models for representative densities and metallicities (\S 2).}}}
\end{figure}

 \autoref{fig:ed_surf} shows the separation surfaces for the cases of the ([O I]/H$\rm{\alpha}$ - [O II]/$\rm{\beta}$ - [O III]/H$\rm{\beta}$; Diagnostic A), ([S II]/H$\rm{\alpha}$ - [O I]/H$\rm{\alpha}$ - [O III]/H$\rm{\beta}$; Diagnostic B) and ([N II]/H$\rm{\alpha}$ - [O I]/H$\rm{\alpha}$ - [O III]/H$\rm{\beta}$; Diagnostic C) diagnostics. The general form of these surfaces is:
\begin{align*} 
F(x, y, z) = \sum_{i=0}^3{a_{ijk}x^iy^jz^k} = 0
\end{align*} 
and the coefficients for each diagnostic are shown in \autoref{table:a_io}. According to these criteria, sources with F(a, b, c) > 0, where a, b and c are the line ratios of the examined source, are shock-excited regions (SNRs). In \autoref{fig:OI_OIII} we present the optimal 2D diagnostic tools for the line ratios ([O I]/H$\rm{\alpha}$ - [O III]/H$\rm{\beta}$; Diagnostic D), ([N II]/H$\rm{\alpha}$ - [O I]/H$\rm{\beta}$; Diagnostic E) and ([O II]/H$\rm{\alpha}$ - [O III]/H$\rm{\beta}$; Diagnostic F). These lines are described by the function:  

\begin{align*} 
G(x, y) = \sum_{i=0}^3{b_{ij}x^iy^j} = 0
\end{align*}
and the respective coefficients are shown in \autoref{table:d_io}. Similarly to the 3D case, sources with G(a, b) > 0, where a and b are the line ratios of the examined source, are considered to be shock-excited regions (SNRs).

\begin{table}
\caption{\small{Coefficients of the decision function (3rd-order polynomial kernel) for the 3D diagnostics.}}
\centering
\begin{threeparttable}
\begin{tabular}{|p{1.5cm}||p{1.5cm}|p{1.5cm}|p{1.5cm}|}
 \hline
 \hfil ijk &\hfil A &\hfil B &\hfil C \\
 \hline
 \hfil 000   &\hfil 3.070    &\hfil 0.303       &\hfil 0.567   \\
 \hfil 010   &\hfil -2.228   &\hfil -1.357      &\hfil -0.862   \\
 \hfil 020   &\hfil -0.554   &\hfil 0.815       &\hfil 0.386   \\
 \hfil 030   &\hfil 0.452    &\hfil 0.696       &\hfil 0.590   \\
 \hfil 001   &\hfil 0.824    &\hfil 1.197       &\hfil 1.400   \\
 \hfil 011   &\hfil 2.248    &\hfil -1.854      &\hfil -2.307   \\
 \hfil 021   &\hfil 0.185    &\hfil 1.356       &\hfil 0.477   \\
 \hfil 002   &\hfil -1.476   &\hfil -1.495      &\hfil -1.213   \\
 \hfil 012   &\hfil -0.771   &\hfil 2.312       &\hfil 1.837   \\
 \hfil 003   &\hfil 0.964    &\hfil 1.874       &\hfil 1.842   \\
 \hfil 100   &\hfil -0.871   &\hfil -1.227      &\hfil -1.752   \\
 \hfil 110   &\hfil 0.932    &\hfil 0.285       &\hfil 2.433   \\
 \hfil 120   &\hfil -1.075   &\hfil 1.766       &\hfil 1.281   \\
 \hfil 101   &\hfil -0.174   &\hfil 1.687       &\hfil -0.428    \\
 \hfil 111   &\hfil -2.650   &\hfil -3.664      &\hfil -3.680   \\
 \hfil 102   &\hfil 1.842    &\hfil -1.329      &\hfil 2.230   \\
 \hfil 200   &\hfil -0.166   &\hfil 0.873       &\hfil 0.027   \\
 \hfil 210   &\hfil 1.134    &\hfil -0.896      &\hfil 1.873   \\
 \hfil 201   &\hfil -0.419   &\hfil 0.403       &\hfil 0.195   \\
 \hfil 300   &\hfil 0.768    &\hfil -0.598      &\hfil -1.082   \\
\hline
\end{tabular}
\begin{tablenotes}
      \item \scriptsize A: [O I]/H$\rm{\alpha}$ - [O II]/H$\rm{\beta}$ - [O III]/H$\rm{\beta}$\\
B: [S II]/H$\rm{\alpha}$ - [O I]/H$\rm{\alpha}$ - [O III]/H$\rm{\beta}$\\
C: [N II]/H$\rm{\alpha}$ - [O I]/H$\rm{\alpha}$ - [O III]/H$\rm{\beta}$
\end{tablenotes}
\end{threeparttable}
\label{table:a_io}
\end{table}

\begin{table}
\caption{\small{Coefficients of the decision function (3rd-order polynomial kernel) for the 2D diagnostics.}}
\begin{threeparttable}
\centering
\begin{tabular}{|p{1cm}||p{1cm}|p{1cm}|p{1cm}|p{1cm}|}
 \hline
 ij & \hfil D & \hfil  E & \hfil  F \\
 \hline
 00   &\hfil  2.710     &\hfil  1.285   &\hfil -2.904 \\
 01   &\hfil  2.096     &\hfil  0.382   &\hfil  3.356\\
 02   &\hfil -1.610     &\hfil  1.263   &\hfil   1.344   \\
 03   &\hfil  0.049     &\hfil  1.162   & \hfil  0.347   \\
 10   &\hfil -0.701     &\hfil -0.0007   & \hfil  4.591  \\
 11   &\hfil -0.887     &\hfil  4.330   &\hfil -1.133 \\
 12   &\hfil  1.067     &\hfil  0.067   &\hfil  -0.616    \\
 20   &\hfil -0.432     &\hfil -0.874   &\hfil 1.952 \\
 21   &\hfil -0.245     &\hfil  4.249   &\hfil 0.416 \\
 30   &\hfil  0.465     &\hfil -2.159   & \hfil  -0.008  \\
\hline
\end{tabular}
\begin{tablenotes}
\item \scriptsize
D: [O I]/H$\rm{\alpha}$ - [O III]/H$\rm{\beta}$\\
E: [N II]/H$\rm{\alpha}$ - [O I]/H$\rm{\alpha}$\\
F: [O II]/H$\rm{\beta}$ - [O III]/H$\rm{\beta}$
\end{tablenotes}
\end{threeparttable}
\label{table:d_io}
\end{table}

\section{Discussion}
\subsection{Effect of Metallicity}
\par{The diagnostics presented in section 3 are based on photo-ionization and shock models for a wide range of metallicities, from $0.25\,Z_{\odot} - 2\,Z_{\odot}$. Since metallicity is directly linked to the strength of the forbidden lines (\citealt{Leonidaki2013}; \citealt{DOdorico_Benvenuti_Sabbadin}), we explore the efficiency of the diagnostics described in \S 3.2 in different metallicity regimes. Hence, we calculate completeness and contamination for the SNRs with subsolar, solar, and supersolar metallicities. These results are shown in \autoref{table:CP_CT_table_sep}. We see that for the cases of subsolar and solar metallicities the diagnostics work quite well while for supersolar metallicities less good. This happens because high-metallicity nebulae have strong temperature gradient (\citealt{Stasinska2004}; \citealt{Stasinska1980}; \citealt{Stasinska1978}) resulting in a wider range of intensities for the oxygen lines. Actually, for supersolar metallicities the intensities of the oxygen lines extend to lower values, compared to solar or subsolar metallicities, and thus shifting the lower-excitation shock excited sources in the HII region locus. }

\begin{table}
\caption{Dependence of the completeness (CP) and contamination (CT) for the different diagnostics on the metallicity}
\begin{tabular}{|p{1.2cm}|p{0.7cm}|p{0.7cm}|p{0.7cm}|p{0.7cm}|p{0.7cm}|p{0.7cm}|}
 \hline
 \multicolumn{7}{|c|}{Subsolar metallicities - $\rm{0.25-0.5\,Z_{\odot}}$ (LMC-SMC metallicities)} \\
 \hline
\hline
Diagnostics:  & \hfil  A & \hfil  B & \hfil  C & \hfil  D & \hfil  E & \hfil  F \\
\hline
 CP   & 0.995 & 0.996 & 0.987 & 0.998 & 0.974 & 0.835 \\
 CT   & 0.078 & 0.102 & 0.081 & 0.075 & 0.144 & 0.317 \\
\hline
\end{tabular}

\begin{tabular}{|p{1.2cm}|p{0.7cm}|p{0.7cm}|p{0.7cm}|p{0.7cm}|p{0.7cm}|p{0.7cm}|}
 \hline
 \multicolumn{7}{|c|}{Solar metallicities} \\
 \hline
\hline
Diagnostics:  & \hfil  A & \hfil B & \hfil  C & \hfil  D & \hfil  E & \hfil  F \\
\hline
 CP   & 0.995 & 0.994 & 0.955 & 0.992 & 0.991 & 0.911    \\
 CT   & 0.017 & 0.023 & 0.017 & 0.016 & 0.032 & 0.087 \\
\hline
\end{tabular}

\begin{tabular}{|p{1.2cm}|p{0.7cm}|p{0.7cm}|p{0.7cm}|p{0.7cm}|p{0.7cm}|p{0.7cm}|}
 \hline
 \multicolumn{7}{|c|}{Supersolar metallicities - $\rm{2\,Z_{\odot}}$} \\
 \hline
\hline
Diagnostics:  & \hfil  A & \hfil  B & \hfil  C & \hfil  D & \hfil  E & \hfil F \\
\hline
 CP   & 0.922 & 0.922 & 0.916 & 0.919 & 0.916 & 0.838    \\
 CT   & 0.155 & 0.185 & 0.158 & 0.139 & 0.247 & 0.460   \\
\hline
\end{tabular}
\label{table:CP_CT_table_sep}
\end{table}

\subsection{Comparison with [S II]/H$\rm{{\alpha}}$ > 0.4 criterion}
\par
\begin{figure*}
\centering
\begin{minipage}{160mm}
\centering
    \includegraphics[width=\textwidth]{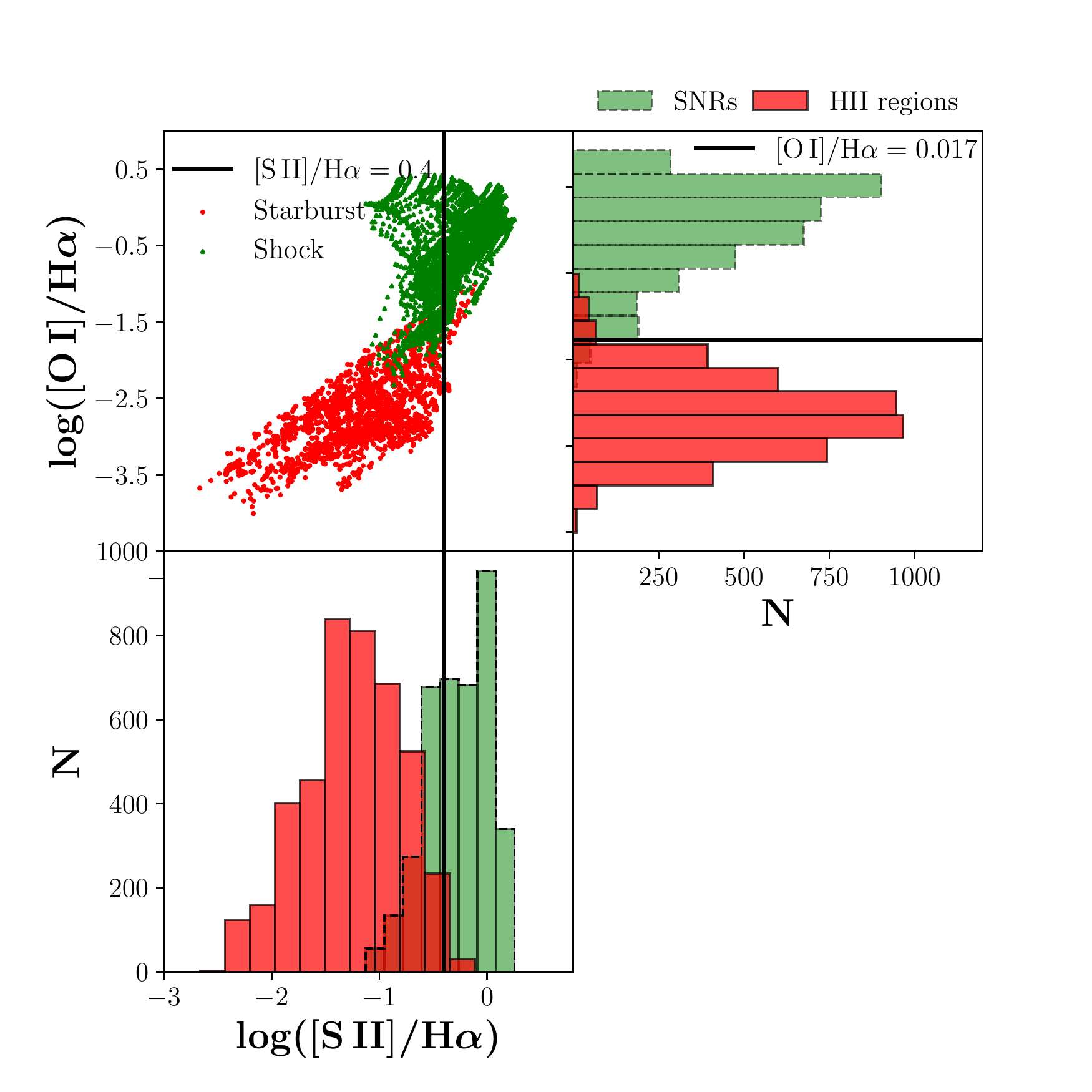}%
\caption{{\label{fig:04_line} \small{Behaviour of the [S II]/H$\rm{\alpha}$ and [O I]/H$\rm{\alpha}$ diagnostics in separating SNRs (green) from HII regions (red). Top left: 2D [S II]/H$\rm{\alpha}$ - [O I]/H$\rm{\alpha}$ diagnostic. The black line indicates the standard [S II]/H$\rm{\alpha} > 0.4$ diagnostic. Top right panel: the histogram of the [O I]/H$\rm{\alpha}$ ratio for SNRs and HII regions along with the 0.017 line (see \S 4.4). This line ratio minimizes the overlap between SNRs and HII regions. Bottom left: histogram of the [S II]/H$\rm{\alpha}$ ratio for SNRs and HII-regions along with the 0.4 line. We see that there are many SNRs with  [S II]/$\rm{H{\alpha} < 0.4}$ that are not identified as such, and a few HII regions that have  [S II]/$\rm{H{\alpha} > 0.4}$ and thus are identified as SNRs using the  [S II]/$\rm{H{\alpha} > 0.4}$ criterion.  In both histograms, N is the number of the points of the  shock and starburst models}.}}
\end{minipage}
\end{figure*}

{The standard diagnostic for identifying SNRs is the [S II]/$\rm{H{\alpha} > 0.4}$ criterion. Here we investigate the efficiency of this diagnostic in the light of the 2D and 3D diagnostics presented in \S 3.2. \autoref{fig:04_line} shows the  [S II]/$\rm{H{\alpha} = 0.4}$ line on the [S II]/H$\rm{\alpha}$ - [O I]/H$\rm{\alpha}$ diagram (top left panel) along with histograms of starburst and shock models for each of the two line ratios ([S II]/H$\rm{\alpha}$ bottom; [O I]/H$\rm{\alpha}$ right). In the case of the  [S II]/H$\rm{\alpha}$ line ratio we also indicate the 0.4 line.  As we can see, there is a significant fraction of shock models with [S II]/H$\rm{\alpha}$ lower than 0.4 (on the left of the 0.4 line) but also a small number of starburst models that have ratios higher than 0.4. This means that by using the  [S II]/$\rm{H{\alpha} > 0.4}$ criterion as selection criterion, we may miss many shock excited sources or identify as SNRs photoionized sources, like HII regions. In \autoref{table:CP_CT_table_all_metal} we give a summary of the completeness and the contamination for the [S II]/$\rm{H{\alpha} > 0.4}$ criterion for all the metallicities together and for the subsolar, solar and supersolar metallicities separately. The effect is more dramatic in the case of subsolar metallicities where we may miss even up to $\sim$70\% of the SNR population. In higher metallicities the effect is weaker but it still may result up to 25\% incompleteness and $\sim$15-20\% contamination by HII-regions.}

\begin{table}
\caption{Completeness - Contamination for the [S II]/H$\rm{\alpha}$ > 0.4 criterion}
\begin{tabular}{|p{1.5cm}|p{1.2cm}|p{1.2cm}|p{1.2cm}|p{1.2cm}|}
\hline
Metallicities:  &Total & Subsolar & Solar & Supersolar \\
\hline
 CP   & 0.658 & 0.317 & 0.682 & 0.764     \\
 CT   & 0.019 & 0.217 & 0.026 & 0.175     \\
\hline
\end{tabular}
\label{table:CP_CT_table_all_metal}
\end{table}

\par {Therefore, the full 2D and 3D diagnostics, give us the possibility to detect up to $\sim$30\% more SNRs than we did up to now.} 
\par {Most importantly the application of the [S II]/$\rm{H{\alpha} > 0.4}$ criterion, leads to a selection effect against slow-shock objects. \autoref{fig:vel_hist} shows a cumulative histogram of the shock velocities of all shock models (i.e any [S II]/H$\rm{\alpha}$ ratio) and those with  [S II]/$\rm{H{\alpha} < 0.4}$. As we can see, SNRs with lower velocities have predominantly  [S II]/$\rm{H{\alpha} < 0.4}$. This selection effect in turn results in a bias against older SNRs which have weaker shocks. In addition, there are SNRs with high velocities that are not detected with the 0.4 criterion. These SNRs are characterized by lower values of magnetic parameters (usually high preshock densities $\rm{100-1000\,cm^{-3}}$ or more rarely low magnetic field $\sim \rm{1 \mu\,G}$). In these cases, the density close to the photoionized zone of the shock becomes high and hence the spontaneous de-excitation of forbidden lines becomes less important (\citealt{Allen2008}), leading to a relatively lower [S II]/H$\rm{\alpha}$ ratio. 
\par {In order to find a 1D diagnostic with which we can recover these slow-shock regions, we constructed histograms for photoionized and shock-excited regions for each line ratio we considered. The line ratio that minimizes the overlap between the two populations is the [O I]/H$\rm{\alpha}$ with 171 (out of 8080) overlapping models (\autoref{fig:04_line} top right panel) and recovers the vast majority of the SNRs that were Missed by the [S II]/H$\rm{\alpha}$ criterion ($\simeq 97\%$ of the total number of the points of the shock models), while keeping the contamination by photoionized regions at a minimum. Of course, the 2D and 3D diagnostics have even higher completeness and lower contamination (c.f. \autoref{table:CP_CT_table}).}

\begin{figure}
 \includegraphics[width=0.5\textwidth]{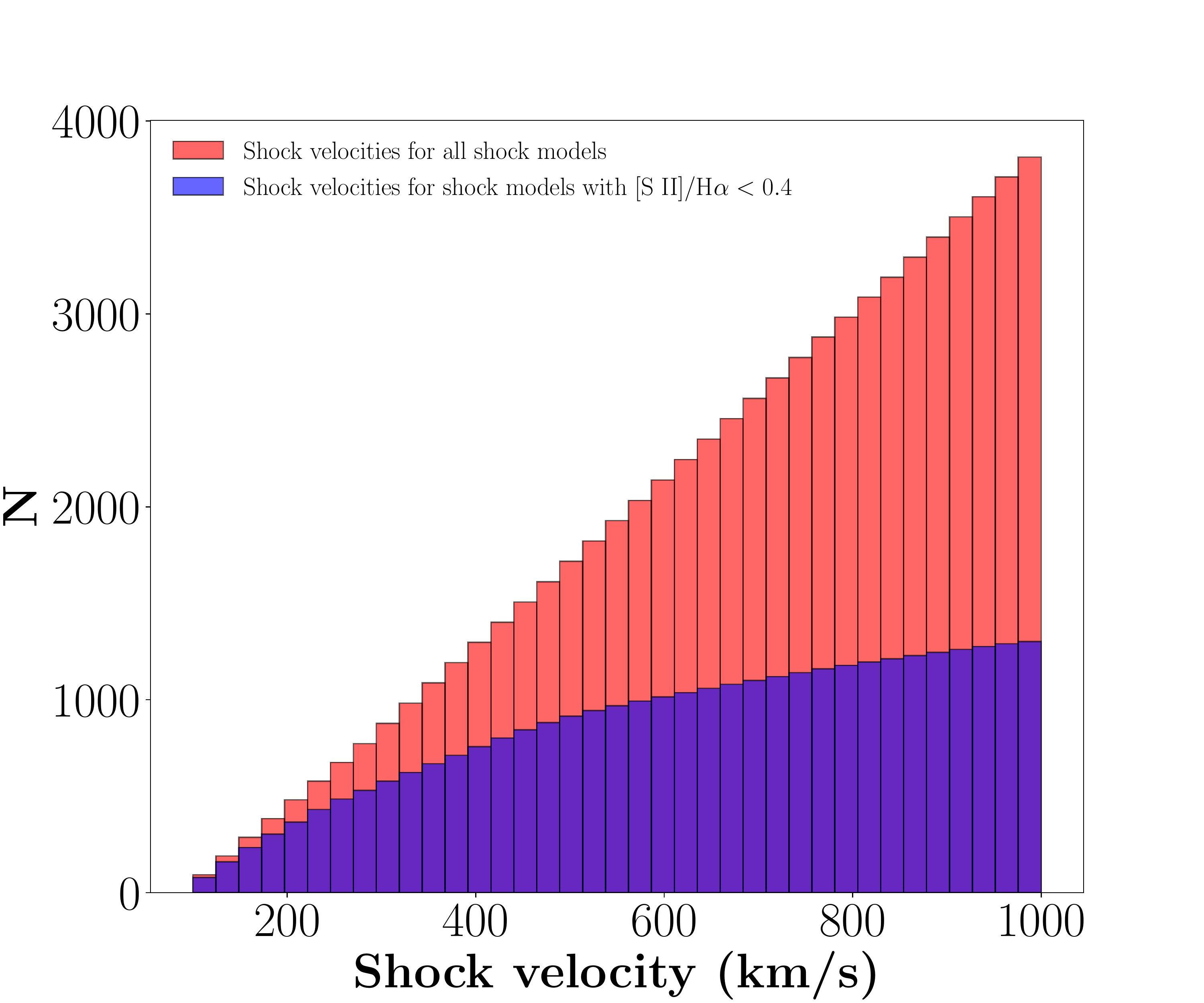}
 \caption{\label{fig:vel_hist} Shock velocity cumulative distribution for shock models of every [S II]/$\rm{H{\alpha} < 0.4}$ ratio (red color) and for shock models with  [S II]/$\rm{H{\alpha} < 0.4}$ (blue color). N is the number of the points of the shock and starburst models.}
\end{figure}

\subsection{Comparison with data}
In order to test the accuracy of the diagnostic tools, we compare our models with observational data. We have divided our data sample into two categories. A sample which refers to Galactic SNRs and SNRs of nearby galaxies (LMC, SMC), the SNR nature of which is confirmed by their morphology and/or their radio properties and consequently we can consider it as a more secure sample. We also consider a second sample which consists of SNRs in more distant galaxies that are identified on the basis of the [S II]/H$\rm{\alpha}$ criterion.
In the same way we use Galactic HII regions and HII regions from the LMC and SMC as a more secure sample and extragalactic HII regions in more distant galaxies as less secure. \autoref{table:data_SNRs} lists individual Galactic sources and the host galaxies for the extragalactic sources, as well as the relevant publications. 
From these studies,  we use objects for which [N II]($\lambda 6583$), [S II]($\lambda\lambda 6716,6731$), [O I]($\lambda 6300$), [O II]($\lambda\lambda 3727, 3729$), [O III]($\lambda 5008$), H$\rm{\alpha}$ and H$\rm{\beta}$ line fluxes are provided.

\begin{table}
\caption{Samples of observational data for SNRs and HII regions}
\noindent\begin{threeparttable}
\noindent
\begin{tabular}{ |lll| }

\hline
\multicolumn{3}{|c|}{$\,\,\,\,\,\,\,\,\,\,\,\,\,\,\,\,\,\,\,\,\,\,\,\,\,\,\,\,\,\,\,\,\,\,\,\,\,\,\,\,\,\,\,\,\,\,\,\,\,\,\,\,\,\,\,\,\,\,\,\,\,\,\,\,\,\,\,\,\,$ SNRs   $\,\,\,\,\,\,\,\,\,\,\,\,\,\,\,\,\,\,\,\,\,\,\,\,\,\,\,\,\,\,\,\,\,\,\,\,\,\,\,\,\,\,\,\,\,\,\,\,\,\,\,\,\,\,\,\,\,\,\,\,\,\,\,\,$              }\\ \hline
\multirow{5}{*}{\citealt{Fesen1985}} & \multirow{1}{*} Galaxy: S147$\_$1$^{*}$, \\ 
& \multirow{1}{*} S147$\_$2$^{*}$, S147$\_$3$^{*}$, S147$\_$4$^{*}$, \\
& \multirow{1}{*} S147$\_$5$^{*}$, ML1$^{\dagger}$, ML2$^{\dagger}$, \\
& \multirow{1}{*} G65.3+5.7$\_$1, G65.3+5.7$\_$2\\ 

\multirow{1}{*}{\citealt{Russel&Dopita}} & SMC, LMC  \\

\multirow{2}{*}{\citealt{Matonick&Fesen}} & \multirow{1}{*}NGC 5204, NGC 5585, NGC 6946\\
                                              & \multirow{1}{*} M81, M101\\

\multirow{2}{*}{\citealt{Leonidaki2013}} & \multirow{1}{*}NGC 2403, NGC 3077, NGC 4214\\
                                              & \multirow{1}{*}NGC 4395, NGC 4449, NGC 5204\\

\multirow{1}{*}{\citealt{Lee2015}} & \multirow{1}{*} M81, M82\\

\multirow{1}{*}{\citealt{Long2019}} & \multirow{1}{*} NGC 6946\\
\hline
\multicolumn{3}{|c|}{$\,\,\,\,\,\,\,\,\,\,\,\,\,\,\,\,\,\,\,\,\,\,\,\,\,\,\,\,\,\,\,\,\,\,\,\,\,\,\,\,\,\,\,\,\,\,\,\,\,\,\,\,\,\,\,\,\,\,\,\,\,\,\,\,\,\,\,$ HII regions $\,\,\,\,\,\,\,\,\,\,\,\,\,\,\,\,\,\,\,\,\,\,\,\,\,\,\,\,\,\,\,\,\,\,\,\,\,\,\,\,\,\,\,\,\,\,\,\,\,\,\,\,\,\,\,\,\,\,\,\,\,\,\,$              }\\ \hline
\multirow{1}{*}{\citealt{zurita&bresolin}} & \multirow{1}{*} M31\\ 

\multirow{1}{*}{\citealt{esteban&bresolin}} & M31, M33  \\

\multirow{1}{*}{\citealt{kwitter&aller}} & \multirow{1}{*} M33\\

\multirow{1}{*}{\citealt{Dufour}} & \multirow{1}{*} LMC\\

\multirow{1}{*}{\citealt{Russel&Dopita}} & \multirow{1}{*} LMC, SMC\\

\multirow{1}{*}{\citealt{Tsamis}} & \multirow{1}{*} LMC, SMC\\

 \multirow{1}{*}{\citealt{vilchez}} & \multirow{1}{*} Galaxy:  S283, S266A, S266B \\

\multirow{1}{*}{\citealt{bresolin2007}} & \multirow{1}{*} M101\\

\multirow{2}{*}{\citealt{castellanos}} & \multirow{1}{*} NGC 628, NGC 925, \\
                                       & \multirow{1}{*} NGC 1637, NGC 1232 \\

\multirow{1}{*}{\citealt{fich}} & \multirow{1}{*} Galaxy: S128, S212\\

\multirow{1}{*}{\citealt{berg}} & \multirow{1}{*} NGC 628\\
\hline
\end{tabular}
\begin{tablenotes}
 \item \scriptsize {$^{*}$\small{\small{These are the positions 1-5 for the SNR S147}.
$^{\dagger}$\small{These are the positions 1 and 2 of Monocerus Loop.}}}\\
\end{tablenotes}
\end{threeparttable}
\label{table:data_SNRs}
\end{table}

\par {We begin with the more secure sample of SNRs which consists of 15 objects. For the  diagnostics B ([S II]/H$\rm{\alpha}$ - [O I]/H$\rm{\alpha}$ - [O III]/H$\rm{\beta}$) and C ([N II]/H$\rm{\alpha}$ - [O I]/H$\rm{\alpha}$ - [O III]/H$\rm{\beta}$) all sources except for two fall in the region of SNRs (given their line ratio uncertainties). In the case of diagnostic D ([O I]/H$\rm{\alpha}$ - [O III]/H$\rm{\beta}$) only one source does not agree (\autoref{fig:comp_data_s_d}) while we have full agreement in the case of diagnostic E ([N II]/H$\rm{\alpha}$ - [O III]/H$\rm{\beta}$) (\autoref{fig:comp_data_s_e}). We note  that the two sources that do not agree with the diagnostics are the same for all the diagnostics and have large uncertainties in the [O I]/H$\rm{\alpha}$ and [O III]/H$\rm{\beta}$ ratios.}

\par {We find between 88.5\%  and 99.2\% agreement between the diagnostics and the prior classification of the less secure sample of SNRs.  The best agreement is for diagnostic E, while the worse agreement  is for diagnostic F.  Figs 9, 10, 11 show these data for the diagnostics D, E and F respectively. Most of the sources that are not found in the expected loci in the diagrams, seem to have very low signal to noise in the [O I] and [O III] lines indicating large uncertainties for [O I]/H$\rm{\alpha}$ and [O III]/H$\rm{\beta}$ line ratios (\citealt{Leonidaki2013}; \citealt{Matonick&Fesen}).}

\par {The more secure sample of HII regions consists of 18 sources. For the diagnostics B and C two out of the 18 sources are found outside of the HII-region locus. One and 2 additional sources are marginally consistent with their respective loci in diagnostics D and E, taking into account the uncertainties when they are available (\autoref{fig:comp_data_s_d}, \autoref{fig:comp_data_s_e}).}

\begin{figure}
\includegraphics[width=0.5\textwidth]{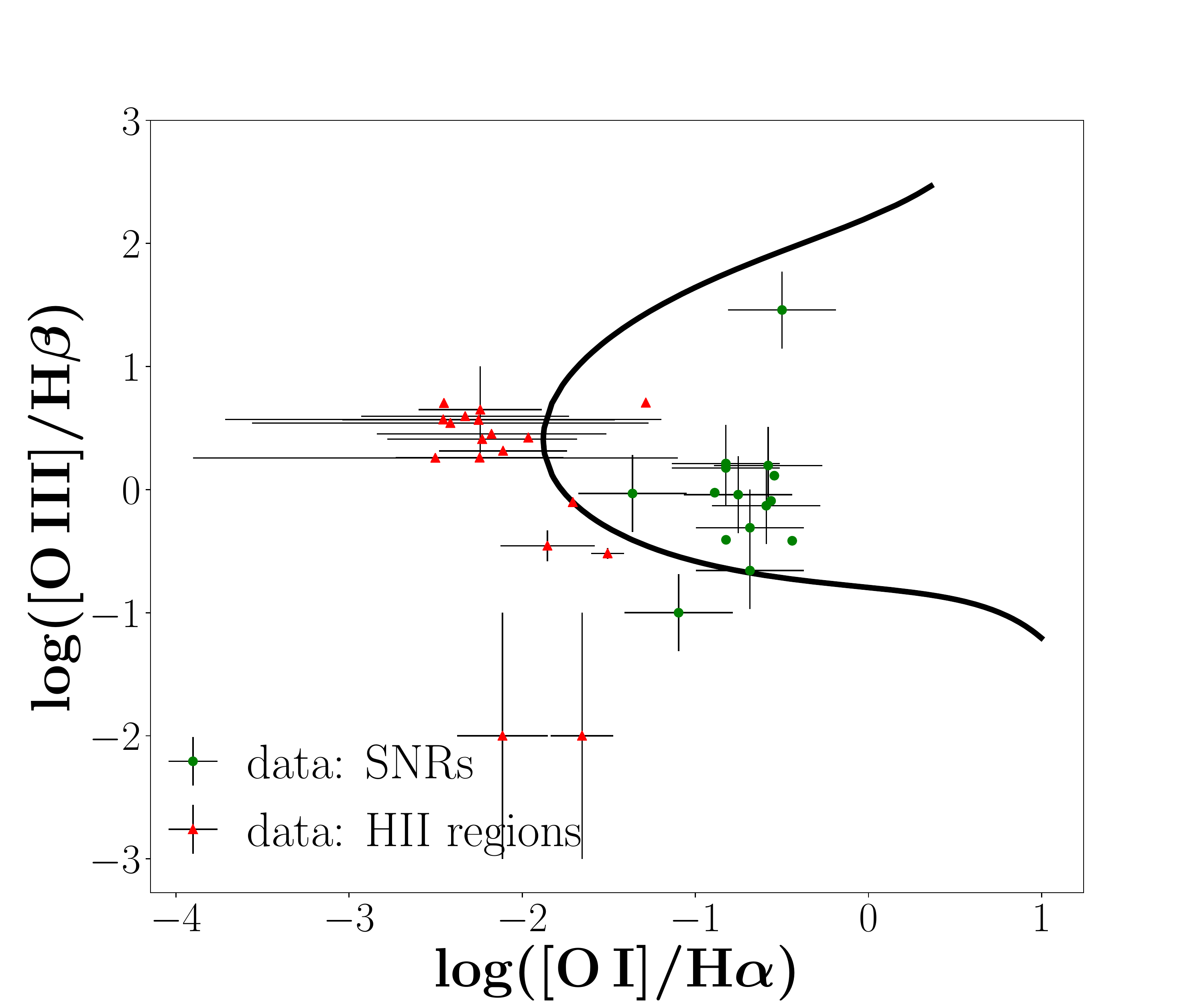}
 \caption{\label{fig:comp_data_s_d} Diagnostic D ([O I]/H$\rm{\alpha}$ - [O III]/H$\rm{\beta}$) for the more secure sample. As we see only 1 out of 15 SNRs is not located in the locus of SNRs and it has large uncertainties in the [O I]/H$\rm{\alpha}$ and [O III]/H$\rm{\beta}$ line ratios. For the HII regions only 1 out of the 18 sources is not found in the HII-region locus. This source, as well as the other sources without errorbars did not have available uncertainties in the respective publications.}
\end{figure}
\begin{figure}
\includegraphics[width=0.5\textwidth]{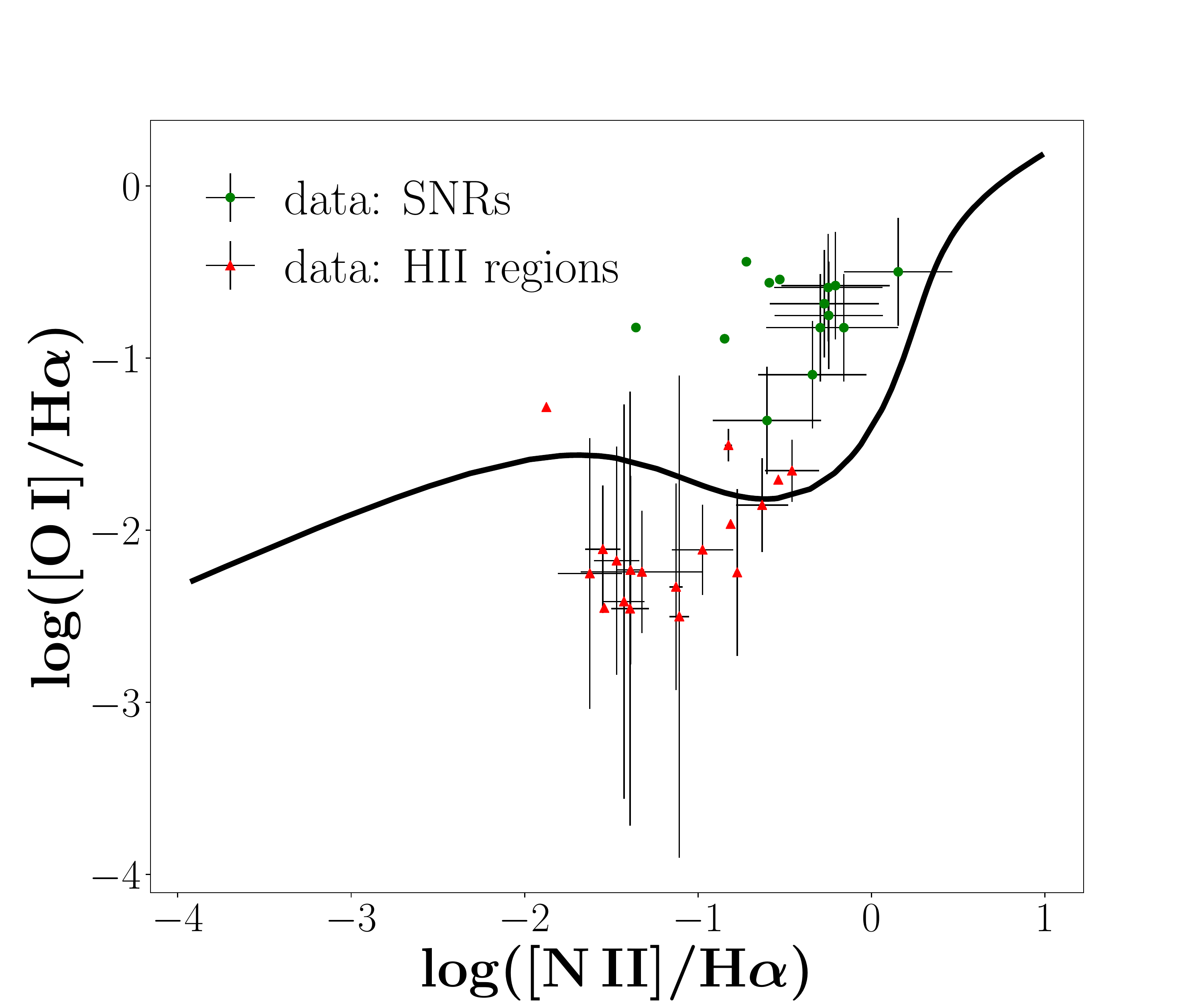}
 \caption{\label{fig:comp_data_s_e} Diagnostic E ([N II]/H$\rm{\alpha}$ - [O I]/H$\rm{\alpha}$) for the more secure sample. As we see all the SNRs are located in the locus defined by the shock regions. For the HII-regions 15 out of 18 are found in the photoionized-regions locus. The errorbars are presented when the line-intensity uncertainties are available in the respective works.}
\end{figure}

\par{For the less secure sample of HII regions we have 100\% agreement for diagnostics A and F (7 out of 7 sources). For the rest of diagnostics (B, C, D and E) $\sim$ 13\% for the sources fall out of the HII-region locus.}

\begin{figure}
\includegraphics[width=0.5\textwidth]{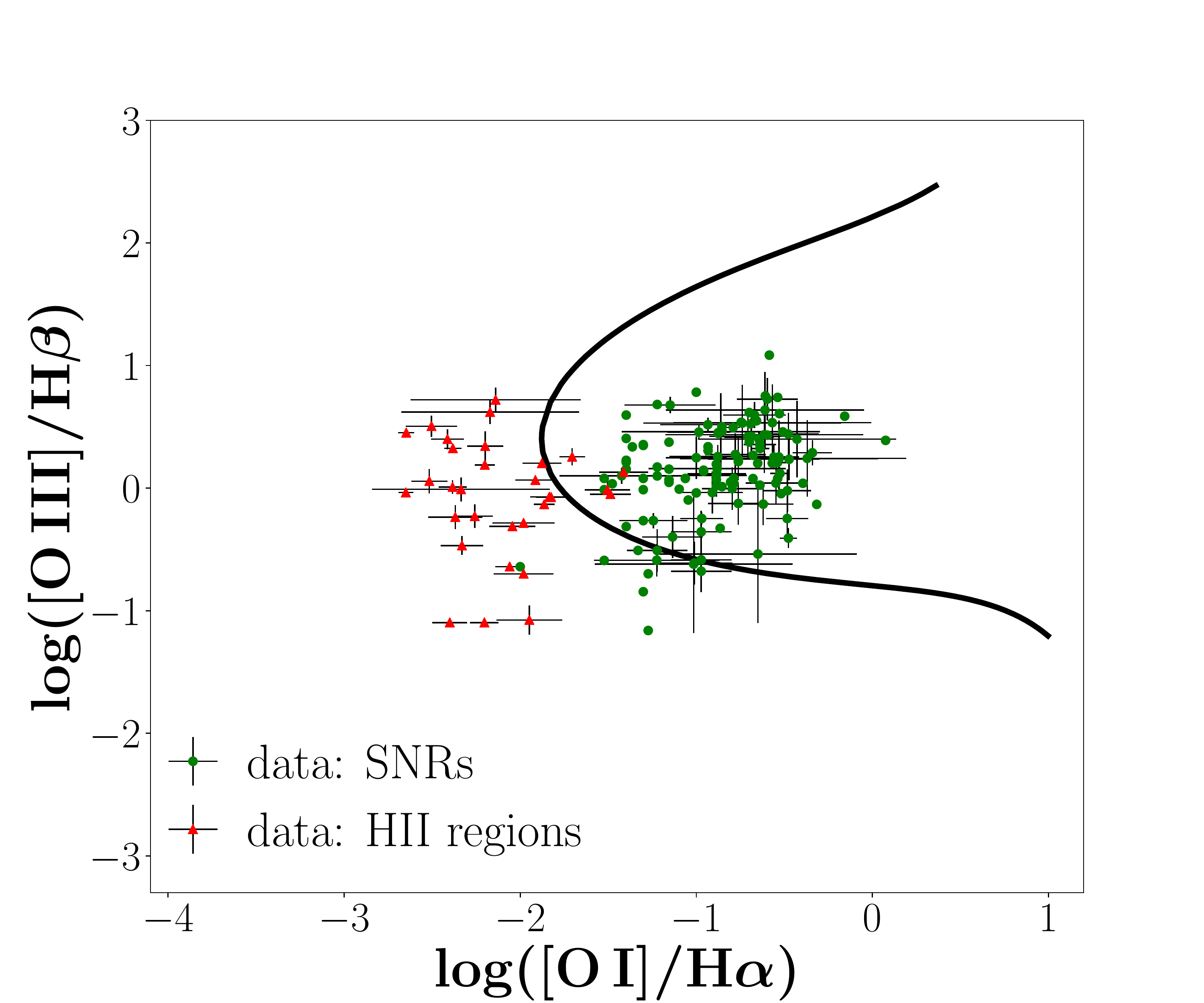}
 \caption{\label{fig:comp_data_d} Diagnostic D ([O I]/H$\rm{\alpha}$ - [O III]/H$\rm{\beta}$) for the less secure sample. As we see 6 out of 127 SNRs are not located in the locus defined by the shock regions. These sources have low signal to noise ratio in the [O I] and [O III] lines indicating large uncertainties in the respective line ratio. The errorbars are presented when the line-intensity uncertainties are available in the respective works.}
\end{figure}
\begin{figure}
\includegraphics[width=0.5\textwidth]{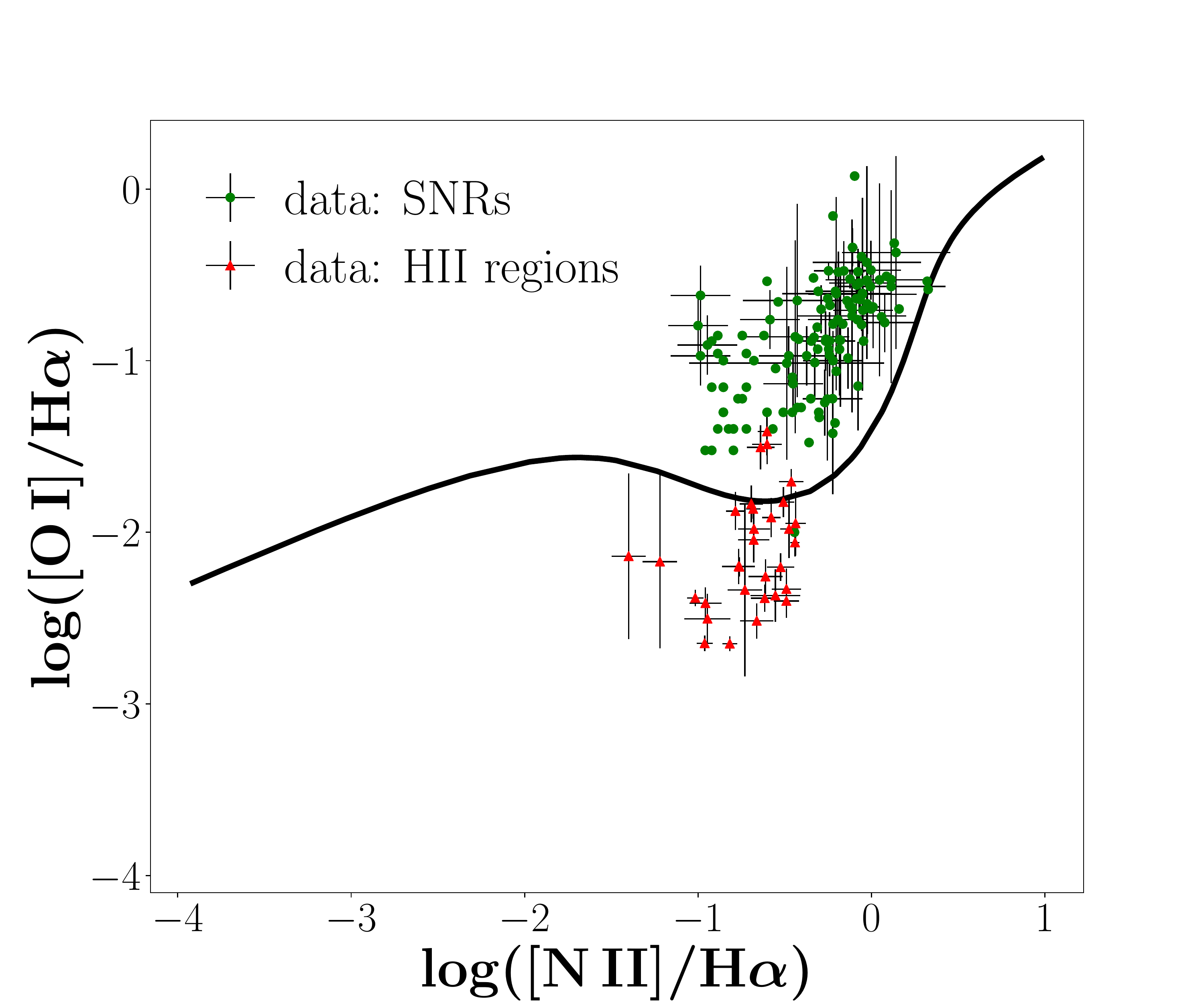}
 \caption{\label{fig:comp_data_e} Diagnostic E ([N II]/H$\rm{\alpha}$ - [O I]/H$\rm{\alpha}$) for the less secure sample. As we see only 1 out of 127 SNRs is not located in the locus defined by the shock regions. This source has low signal to noise ratio in the [O I] line indicating large uncertainties in the respective line ratio. The errorbars are presented when the line-intensity uncertainties are available in the respective works.}
\end{figure}

\begin{figure}
\includegraphics[width=0.5\textwidth]{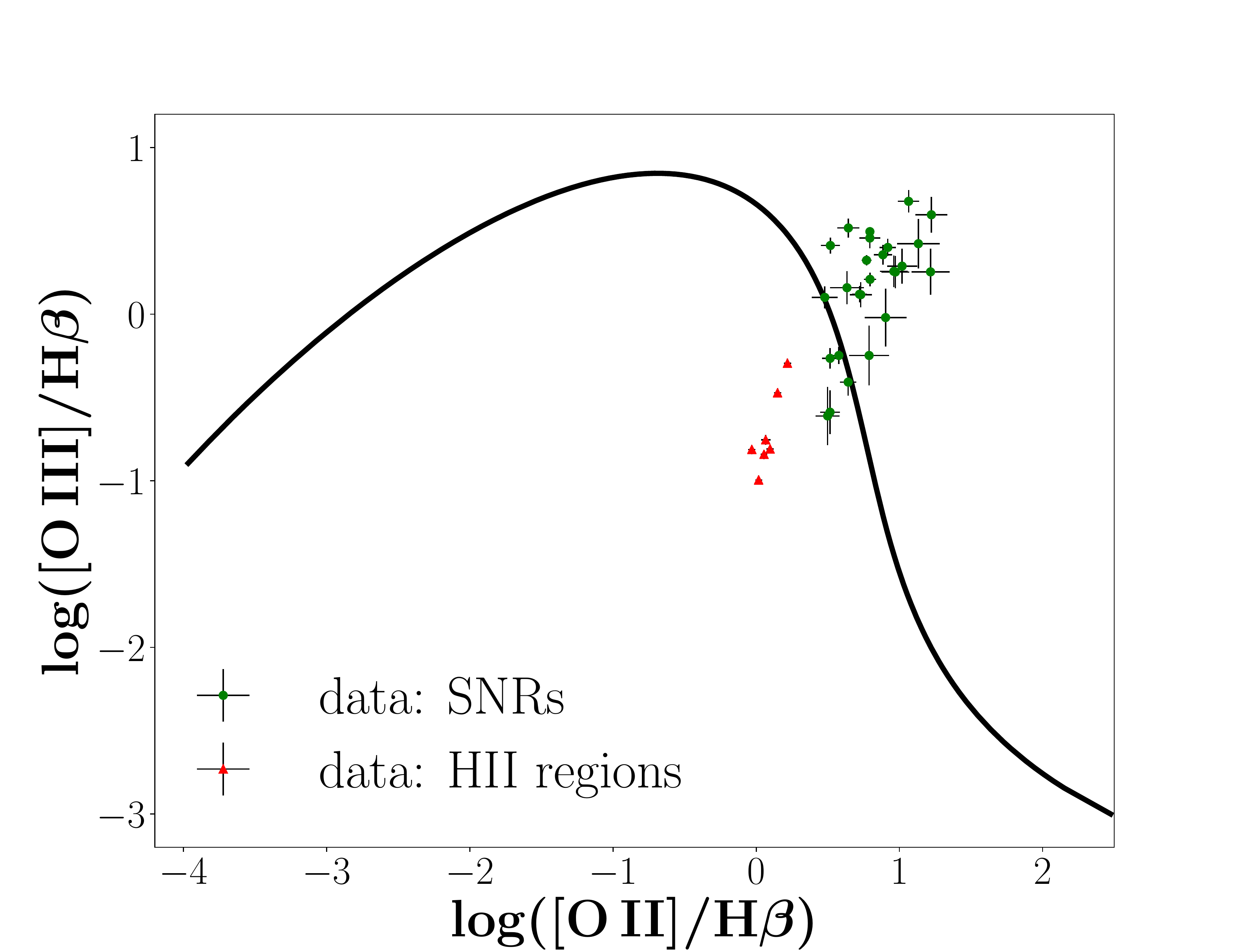}
 \caption{\label{fig:comp_data_f} Diagnostic F ([O II]/H$\rm{\beta}$ - [O III]/H$\rm{\beta}$) for the less secure sample. In this case 23 out of 26 SNRs fall in the SNR locus, while all the HII regions are located in the HII-region locus.}
\end{figure}

\par{In summary, we find very good agreement between the diagnostics  and the morphologically selected SNRs (\autoref{fig:comp_data_s_d}, \autoref{fig:comp_data_s_e}). In addition, even though we do not expect 100\% agreement between our diagnostics and the less secure sample, since they have been selected based on the [S II]/H$\rm{\alpha}$ > 0.4 criterion,  we find that they agree very well.}
\par{The same holds in the case of HII regions. Observed HII regions are clearly separated from the SNRs resulting in minimal or no contamination of the SNR population by HII regions.}}  
\par {Furthermore, when emission-line uncertainties are available, we can account for those and derive the probability of a source to belong to shock-excited or photoionized region locus (indicating SNR and HII regions respectively; e.g. by means of Monte Carlo sampling, e.g. \citealt{Maragkoudakis}).

\par {One complication in the identification of SNRs on the basis of their line ratios is objects that are embedded in HII regions. In this case, the generally weaker higher excitation lines}of the HII region would shift the location of the SNRs away from their locus on the diagnostic diagram. Determining a diagnostic that accounts for the contamination by the surrounding HII regions, is beyond the scope of this  paper and will be presented in a forthcoming paper.}

\par As we can see, in some cases there are contradicting classifications. Sources that have been classified as SNRs according to a specific diagnostic, are classified as HII regions using other diagnostic. For example,  While almost all of the observed SNRs are classified as shock-excited regions using diagnostic E (Figs 8, 10) a few of them are classified as photoionized (HII) regions based on diagnostic D (Figs 7, 9). 
This is expected since these diagnostics are simply projections of the multidimensional manifold of the distribution of the shock-excited (SNR) and photoionized (HII regions) in the parameter space defined by the spectral line ratios (e.g. \citealt{Stampoulis2019}). Obviously higher dimensionality diagnostics (e.g. diagnostics A, B, and C) offer much better consistency since \textit{all} available lines are used simultaneously. However, this comes at the cost of requiring measurements of multiple lines, some of which are rather weak.

\subsection{Possible biases and comparison with other object classes}
The diagnostics presented here are based on the comparison of the ratio between different forbidden lines and their corresponding (closest) Balmer lines. They are an extension of the commonly used [S II]/H$\alpha$ diagnostic to include other diagnostically powerful line ratios combined with a quantitative definition of the diagnostic. Since they also employ forbidden lines they suffer from the same bias against Balmer dominated SNR inherent in the traditional [S II]/H$\alpha$ diagnostic. This class of SNRs is characterized by weak or absent, forbidden lines and they are traditionally recognized on the basis of their strong and broad Balmer lines (e.g.   \citealt{Heng2010}). However, this is not a strong bias for studies of the overall population of SNRs, since Balmer dominated SNRs are only a small subset of the optically emitting SNR population.
\par Other types of objects that also produce high excitation lines are planetary nebulae and Herbig-Haro objects. However planetary nebulae are characterized by strong [O III] emission and weak [S II], [O I] or [N II] emission which would discriminate them from the locus of SNRs in our diagnostic diagrams and place them in the high-excitation end of the HII-region locus (e.g. \citealt{baldwin1981}, \citealt{sabbadin1977}).
\par On the other hand although Herbig-Haro objects are excited by the shock of the jets of young stellar objects, their total luminosity ($\sim 0.1\rm{L\odot}$ ; e.g. \citealt{Riaz2017}) renders them  unobservable in SNR surveys of nearby galaxies. In our Galaxy they can be easily discriminated from SNRs on the basis of their morphology.

\subsection{Suggested tool for photometric selection of SNRs}

\par{The results presented in section \S 3.2 show that the ideal diagnostic combines the [O I], [O III], [O II], or [S II] forbidden lines along with their corresponding Balmer lines (H$\rm{\alpha}$ and H$\rm{\beta}$). However this requires observations in five narrow-band filters which greatly increases the required telescope time.} 
\par{In \autoref{fig:04_line}, we compare the distributions of the [S II]/H$\rm{\alpha}$ and [O I]/H$\rm{\alpha}$ line ratios for the SNRs and HII regions. As is clearly seen, the [O I]/H$\rm{\alpha}$ line ratio separates more effectively the HII regions from SNRs (see also \citealt{Lee2015}, \citealt{Fesen1985}). This is because the [O I] line is produced in the interface between the photoionized HII region and the surrounding material. In HII region, this interface tends to be quite narrow, since almost all the oxygen is ionized (\citealt{Evans_Dopita1985}),  resulting in weaker [O I] emission. On the other hand, in SNRs because of the different excitation mechanism and the presence of the photoionizing precursor, the size of this region is wider resulting in stronger [O I] emission. Consequently, H$\rm{\alpha}$ and [O I](6300) narrow-band filters can be used for SNR  selection. In this case SNR candidates are sources with [O I]/H$\rm{\alpha}$ ratio higher than 0.017 (or log{([O I]/$\rm{H\rm{\alpha}) > -1.76}$) (\autoref{fig:04_line} top right). The completeness for SNR selection using this diagnostic is 97.2\% and the contamination only 2.4\%.}

\section{Conclusions}

In this paper we presented theory-driven line-ratio diagnostics for the identification of SNRs. These diagnostics are very promising in reducing the bias against lower excitation SNRs in comparison to the traditional [S II]/H$\rm{\alpha}$ diagnostic and they can increase the number of identified SNRs by $\sim{30\%}$ at least. We explore six line-ratios combined in 3D or 2D diagnostics involving [O I], [O II], [O III], [S II] and [N II] lines and their corresponding H$\rm{\alpha}$ and H$\rm{\beta}$ lines. We find that  the best 3D and 2D diagnostics in terms of their completeness and low contamination  by HII regions  are [O I]/H$\rm{\alpha}$ - [O II]/H$\rm{\beta}$ - [O III]/H$\rm{\beta}$ and  [O I]/H$\rm{\alpha}$ -[O III]/H$\rm{\beta}$ respectively. 
We also find that the [O I]/H$\rm{\alpha}$ diagnostic is very efficient for selecting SNRs, in agreement with previous reports. Here we define the selection criterion ([O I]/$\rm{H{\alpha} > 0.017}$) and we quantify its completeness (97.2\%) and contamination (2.4\%).  This efficiency is significantly higher than the one of 65.8\% for the [S II]/H$\rm{\alpha}$ > 0.4 diagnostic that has been used up to now.

%

This work has been based on MAPPINGS III shock and starburst models. The use of other shock and starburst models would give probably different diagnostics (different separating lines and surfaces), however, the capabilities  of the different line-ratio combinations in distinguishing SNRs from HII regions should be the same or at least very similar.

\section*{Acknowledgements}
We thank the anonymous referee for the helpful comments that helped to improve the clarity of the paper. We acknowledge funding from the European Research Council under the European Union's Seventh Framework Programme (FP/2007-2013)/ERC Grant Agreement n. 617001. This project has received funding from the European Union's Horizon 2020 research and innovation programme  under the Marie Sklodowska-Curie RISE action,  grant agreement No 691164 (ASTROSTAT). We also thank Jeff Andrews for organizing the SMAC (Statistical methods for Astrophysics in Crete) seminar and for helpful discussions about SVM, and Paul Sell for his important help on the construction of 3D animations.





\clearpage  
\begin{thebibliography}{99}

\bibitem[\protect\citeauthoryear{Allen et al.}{2008}]{Allen2008}
Allen, M. ~G., Groves, B. ~A., Dopita, M. ~A., Sutherland, R. ~S., \& Kewley, L. ~J.\ 2008, \apjs, 178, 20 
\bibitem[\protect\citeauthoryear{Baldwin et al.}{1981}]{baldwin1981}
Baldwin, J.~A., Phillips, M.~M., \& Terlevich, R.\ 1981, \pasp, 93, 5 
\bibitem[\protect\citeauthoryear{Berg et al.}{2015}]{berg}
Berg, D. ~A., Skillman, E. ~D., Croxall, K. ~V., et al.\ 2015, \apj, 806, 16 
\bibitem[\protect\citeauthoryear{Binette et al.}{1985}]{Binette1985}
Binette, L., Dopita, M.~A., \& Tuohy, I.~R.\ 1985, \apj, 297, 476 
\bibitem[\protect\citeauthoryear{Blair et al.}{2013}]{BlairWinklerLong2013}
Blair, W. ~P., Winkler, P. ~F., \& Long, K. ~S.\ 2013, \apjs, 207, 40
\bibitem[\protect\citeauthoryear{Blair et al.}{2012}]{BlairWinklerLong2012}
Blair, W. ~P., Winkler, P. ~F., \& Long, K. ~S.\ 2012, \apjs, 203, 8
\bibitem[\protect\citeauthoryear{Blair \& Long}{1997}]{Blair1997}
Blair, W. ~P., \& Long, K. ~S.\ 1997, \apjs, 108, 261
\bibitem[\protect\citeauthoryear{Blair et al.}{1982}]{Blair1982}
Blair, W. ~P., Kirshner, R. ~P., \& Chevalier, R. ~A.\ 1982, \apj, 254, 50 
\bibitem[\protect\citeauthoryear{Boumis et al.}{2009}]{Boumis2009}
Boumis, P., Xilouris, E. ~M., Alikakos, J., et al.\ 2009, \aap, 499, 789 
\bibitem[\protect\citeauthoryear{Bresolin}{2007}]{bresolin2007}
Bresolin, F.\ 2007, \apj, 656, 186
\bibitem[\protect\citeauthoryear{Castellanos et al.}{2002}]{castellanos}
Castellanos, M., D{\'{\i}}az, A. ~I., \& Terlevich, E.\ 2002, \mnras, 329, 315 
\bibitem[\protect\citeauthoryear{D'Odorico et al.}{1978}]{DOdorico_Benvenuti_Sabbadin}
Dodorico, S., Benvenuti, P., \& Sabbadin, F.\ 1978, \aap, 63, 63 
\bibitem[\protect\citeauthoryear{Daltabuit et al.}{1976}]{Daltabuit_DOdorico_Sabbadin}
Daltabuit, E., Dodorico, S., \& Sabbadin, F.\ 1976, \aap, 52, 93 
\bibitem[\protect\citeauthoryear{Dennefeld \& Kunth}{1981}]{Dennefeld}
Dennefeld, M., \& Kunth, D.\ 1981, \aj, 86, 989 
\bibitem[\protect\citeauthoryear{Dopita et al.}{2010}]{Dopita2010}
Dopita, M. ~A., Blair, W. ~P., Long, K. ~S., et al.\ 2010, \apj, 710, 964 
\bibitem[\protect\citeauthoryear{Dopita et al.}{2005}]{Dopita2005}
Dopita, M. ~A., Groves, B. ~A., Fischera, J., et al.\ 2005, \apj, 619, 755 
\bibitem[\protect\citeauthoryear{Dopita et al.}{2002}]{Dopita_Groves}
Dopita, M. ~A., Groves, B. ~A., Sutherland, R. ~S., Binette, L., \& Cecil, G.\ 2002, \apj, 572, 753 
\bibitem[\protect\citeauthoryear{Dufour}{1975}]{Dufour}
Dufour, R. ~J.\ 1975, \apj, 195, 315 
\bibitem[\protect\citeauthoryear{Esteban et al.}{2009}]{esteban&bresolin}
Esteban, C., Bresolin, F., Peimbert, M., et al.\ 2009, \apj, 700, 654 
\bibitem[\protect\citeauthoryear{Evans \& Dopita}{1985}]{Evans_Dopita1985}
Evans, I. ~N., \& Dopita, M. ~A.\ 1985, \apjs, 58, 125 
\bibitem[\protect\citeauthoryear{Fesen et al.}{1985}]{Fesen1985}
Fesen, R. ~A., Blair, W. ~P., \& Kirshner, R. ~P.\ 1985, \apj, 292, 29 
\bibitem[\protect\citeauthoryear{Fich \& Silkey}{1991}]{fich}
Fich, M., \& Silkey, M.\ 1991, \apj, 366, 107 
\bibitem[\protect\citeauthoryear{Fioc \& Rocca-Volmerange}{1997}]{PEGASE2}
Fioc, M., \& Rocca-Volmerange, B.\ 1997, \aap, 326, 950 
\bibitem[\protect\citeauthoryear{Green}{2017}]{Green2017}
Green, D. ~A.\ 2017, VizieR Online Data Catalog, 7278, 
\bibitem[\protect\citeauthoryear{Groves \& Allen}{2013}]{ITERA2}
Groves, B. ~A., Allen, M. ~G.\ 2013, ITERA: Tool for Emission-line Ratio Analysis. Astrophysics Source Code Library, ascl:1307.012
\bibitem[\protect\citeauthoryear{Groves \& Allen}{2010}]{ITERA1}
Groves, B. ~A., \& Allen, M. ~G.\ 2010, \na, 15, 614 
\bibitem[\protect\citeauthoryear{Groves et al.}{2004}]{Groves2004}
Groves, B. ~A., Dopita, M. ~A., \& Sutherland, R. ~S.\ 2004, \apjs, 153, 9 
\bibitem[\protect\citeauthoryear{Heng}{2010}]{Heng2010} 
Heng, K.\ 2010, \pasa, 27, 23  
\bibitem[\protect\citeauthoryear{Ivezi\'{c} et al.}{2014}]{stat_book}
Ivezi\'{c}, $\check{Z}$, Connolly, A. ~J., VanderPlas, J. ~T. \& Gray, A.\ 2014, "Statistics, Data Mining and Machine Learning in Astronomy", Princeton University Press
\bibitem[\protect\citeauthoryear{Kewley et al.}{2001}]{kewly2000}
Kewley, L. ~J., Dopita, M. ~A., Sutherland, R. ~S., Heisler, C. ~A., \& Trevena, J.\ 2001, \apj, 556, 121
\bibitem[\protect\citeauthoryear{Kwitter \& Aller}{1980}]{kwitter&aller}
Kwitter, K. ~B. \& Aller, L. ~H.\ 1980, \mnras, 195, 939
\bibitem[\protect\citeauthoryear{Lee \& Lee}{2015}]{Lee2015}
Lee, M. ~G., Sohn, J., Lee, J. ~H., et al.\ 2015, \apj, 804, 63 
\bibitem[\protect\citeauthoryear{Lee \& Lee}{2014}]{Lee2014}
Lee, J. ~H., \& Lee, M. ~G.\ 2014, \apj, 786, 130 
\bibitem[\protect\citeauthoryear{Leitherer et al.}{1999}]{Starburst99}
Leitherer, C., Schaerer, D., Goldader, J. ~D., et al.\ 1999, \apjs, 123, 3 
\bibitem[\protect\citeauthoryear{Leonidaki et al.}{2013}]{Leonidaki2013}
Leonidaki, I., Boumis, P., \& Zezas, A.\ 2013, \mnras, 429, 189
\bibitem[\protect\citeauthoryear{Leonidaki et al.}{2010}]{Leonidaki2010}
Leonidaki, I., Zezas, A., \& Boumis, P.\ 2010, \apj, 725, 842
\bibitem[\protect\citeauthoryear{Levesque et al.}{2010}]{levesque2010}
 Levesque, E. ~M., Kewley, L. ~J., \& Larson, K. ~L.\ 2010, \aj, 139, 712 
\bibitem[\protect\citeauthoryear{Long et al.}{2019}]{Long2019}
Long, K.~S., Winkler, P.~F., \& Blair, W.~P.\ 2019, \apj, 875, 85 
\bibitem[\protect\citeauthoryear{Long et al.}{2018}]{Long2018}
Long, K.~S., Blair, W.~P., Milisavljevic, D., Raymond, J.~C., \& Winkler, P.~F.\ 2018, \apj, 855, 140 
\bibitem[\protect\citeauthoryear{Maragkoudakis et al.}{2018}]{Maragkoudakis}
Maragkoudakis, A., Zezas, A., Ashby, M.~L.~N., \& Willner, S.~P.\ 2018, \mnras, 475, 1485 
\bibitem[\protect\citeauthoryear{Mathewson \& Clarke}{1973}]{Mathewson&Clarke}
Mathewson, D.~S., \& Clarke, J.~N.\ 1973, \apj, 180, 725
\bibitem[\protect\citeauthoryear{Matonick \& Fesen}{1997}]{Matonick&Fesen}
Matonick, D.~M., \& Fesen, R.~A.\ 1997, \apjs, 112, 49
\bibitem[\protect\citeauthoryear{Milisavljevic \& Fesen}{2013}]{Dan2013}
Milisavljevic, D., \& Fesen, R. ~A. 2013, \apj, 772, 134
\bibitem[\protect\citeauthoryear{Riaz et al.}{2017}]{Riaz2017}
Riaz, B., Brice{\~n}o, C., Whelan, E.~T., \& Heathcote, S.\ 2017, \apj, 844, 47 
\bibitem[\protect\citeauthoryear{Russel \& Dopita}{1990}]{Russel&Dopita}
Russell, S.~C., \& Dopita, M.~A.\ 1990, \apjs, 74, 93 
\bibitem[\protect\citeauthoryear{Sabbadin et al.}{1977}]{sabbadin1977}
Sabbadin, F., Minello, S., \& Bianchini, A.\ 1977, \aap, 60, 147 
\bibitem[\protect\citeauthoryear{de Souza et al.}{2017}]{Souza2017}
de Souza, R. ~S.,   Dantas, M. ~L. ~L., Costa-Duarte, M. ~V., Feigelson, E. ~D,. Killedar,  M.  Lablanche, P.-Y., Vilalta,  R.,  Krone-Martins, A. ,Beck, R., Gieseke, F. \ 2017 \mnras, 472, 2808
\bibitem[\protect\citeauthoryear{Stampoulis et al.}{2019}]{Stampoulis2019}
Stampoulis, V., van Dyk, D.~A., Kashyap, V.~L., \& Zezas, A.\ 2019, \mnras, 485, 1085
\bibitem[\protect\citeauthoryear{Stasi\'{n}ka}{2005}]{Stasinska2004}
Stasi{\'n}ska, G.\ 2005, \aap, 434, 507  
\bibitem[\protect\citeauthoryear{Stasi\'{n}ka}{1980}]{Stasinska1980}
Stasi{\'n}ska, G.\ 1980, \aap, 85, 359 
\bibitem[\protect\citeauthoryear{Stasi\'{n}ka}{1978b}]{Stasinska1978}
Stasi\'{n}ka, G.\ 1978b, \aaps, 32, 429 
\bibitem[\protect\citeauthoryear{Sutherland \& Dopita}{1993}]{Sutherland_Dopita}
Sutherland, R.~S., \& Dopita, M.~A.\ 1993, \apjs, 88, 253 
\bibitem[\protect\citeauthoryear{Tsamis et al.}{2003}]{Tsamis}
Tsamis, Y.~G., Barlow, M.~J., Liu, X.-W., Danziger, I.~J., \& Storey, P.~J.\ 2003, \mnras, 338, 687
\bibitem[\protect\citeauthoryear{V\'{a}zquez \& Leitherer}{2005}]{vazquez_leitherer2005}
V{\'a}zquez, G.~A., \& Leitherer, C.\ 2005, \apj, 621, 695 
\bibitem[\protect\citeauthoryear{V\'{i}lchez \& Esteban}{1996}]{vilchez}
Vilchez, J.~M. , \& Esteban, C. \ 1996, \mnras, 280, 720
\bibitem[\protect\citeauthoryear{Vogt et al.}{2014}]{Vogt2014}
Vogt, F.~P.~A., Dopita, M.~A., Kewley, L.~J., et al.\ 2014, \apj, 793, 127 
\bibitem[\protect\citeauthoryear{Vu\v{c}eti\'{c} et al.}{2015}]{Vucetic2015}
Vu{\v c}eti{\'c}, M.~M., Arbutina, B., \& Uro{\v s}evi{\'c}, D.\ 2015, \mnras, 446, 943 
\bibitem[\protect\citeauthoryear{Zurita \& Bresolin}{2012}]{zurita&bresolin}
Zurita, A., \& Bresolin, F.\ 2012, \mnras, 427, 1463 

\end{thebibliography}


\clearpage 


%
\appendix
\section{Other Diagnostics}

\begin{table}
\begin{minipage}{\textwidth}
\caption{Completeness, contamination and $\gamma$ parameter for 3D and 2D diagnostics}
\begin{tabular}{|l|c|c|c|}
\hline
Diagnostics  & Completeness & Contamination & $\gamma$\\
\hline
$ \rm{[N\,II]/H\alpha}-[O\,I]/H\rm{\alpha}-[O\,II]/H\rm{\beta}$  & 0.984 & 0.019  &  1.0  \\
$ \rm{[N\,II]/H\alpha}-[O\,II]/H\rm{\beta}-[O\,III]/H\rm{\beta}$  & 0.937 & 0.034 & 1.0 \\
$ \rm{[N\,II]/H\alpha}-[S\,II]/H\rm{\alpha}-[O\,I]/H\rm{\alpha}$ & 0.984 & 0.024 & 0.2   \\
$ \rm{[N\,II]/H\alpha}-[S\,II]/H\rm{\alpha}-[O\,II]/H\rm{\beta}$ & 0.926 & 0.086 &  0.2 \\
$ \rm{[N\,II]/H\alpha}-[S\,II]/H\rm{\alpha}-[O\,III]/H\rm{\beta}$ & 0.973 & 0.015 & 1.0   \\
$ \rm{[S\,II]/H\alpha}-[O\,I]/H\rm{\alpha}-[O\,II]/H\rm{\beta}$ & 0.988 & 0.025 &  1.0   \\
$ \rm{[S\,II]/H\alpha}-[O\,II]/H\rm{\beta}-[O\,III]/H\rm{\beta}$   & 0.965 & 0.012 & 1.0  \\
\hline
$ \rm{[N\,II]/H\alpha}-[O\,II]/H\rm{\beta}$  & 0.827 & 0.102 &  all (0.2 - 1.0)\\
$ \rm{[N\,II]/H\alpha}-[O\,III]/H\rm{\beta}$  & 0.898 & 0.027 &  all (0.2 - 1.0)  \\
$ \rm{[O\,I]/H\alpha}-[O\,II]/H\rm{\beta}$ & 0.985 & 0.022 &  1.0 \\
$ \rm{[S\,II]/H\alpha}-[O\,I]/H\rm{\alpha}$ & 0.983 & 0.026 &  1.0   \\
$ \rm{[S\,II]/H\alpha}-[O\,III]/H\rm{\beta}$ & 0.959 & 0.013  &  0.4\\
$ \rm{[S\,II]/H\alpha}-[O\,II]/H\rm{\beta}$ & 0.941 & 0.142  & 0.4 \\
$ \rm{[S\,II]/H\alpha}-[N\,II]/H\rm{\alpha}$ & 0.926 & 0.104  & 0.2 \\
\hline \\

\label{table:CP_CT_rest_3d}
\end{tabular}
\end{minipage}
\end{table}

Here we present the rest of the diagnostics we examined and are not included in \S 3. 
In \autoref{table:CP_CT_rest_3d} we show the completeness, the contamination and the $\gamma$ perameters of the decision function of the 3D and 2D diagnostics. Figures \ref{fig:NII_OI_OII} - \ref{fig:SII_OII_line} show the separating surfaces and lines for these 3D (\ref{fig:NII_OI_OII}-\ref{fig:SII_OII_OIII}) and 2D (\ref{fig:SII_NII_line}-\ref{fig:SII_OII_line}) diagnostics. Animations that show
the rotation of the surfaces are available in the on-line version. As discussed in \S 3.2 the surfaces are described by the function $F(x, y, z) = \sum_{i=0}^3{a_{ijk}x^iy^jz^k} = 0 $ and the 2D lines by $G(x, y) = \sum_{i=0}^3{b_{ij}x^iy^j} = 0 $. The coefficients of these functions are given in \autoref{table:3d_function} and \autoref{table:2d_function} respectively. For both cases, sources with F > 0 or G > 0 are SNRs.

\par{In all cases, polynomial kernels seem to work better, except for the line-ratio combinations $\rm{[N\,II]/H\alpha-[O\,II]/H\beta}$ and $\rm{[N\,II]/H\alpha-[O\,III]/H\beta}$ for which linear kernel gives better results.}

\begin{figure*}
\begin{minipage}[b]{0.48\textwidth}
    \includegraphics[width=\textwidth]{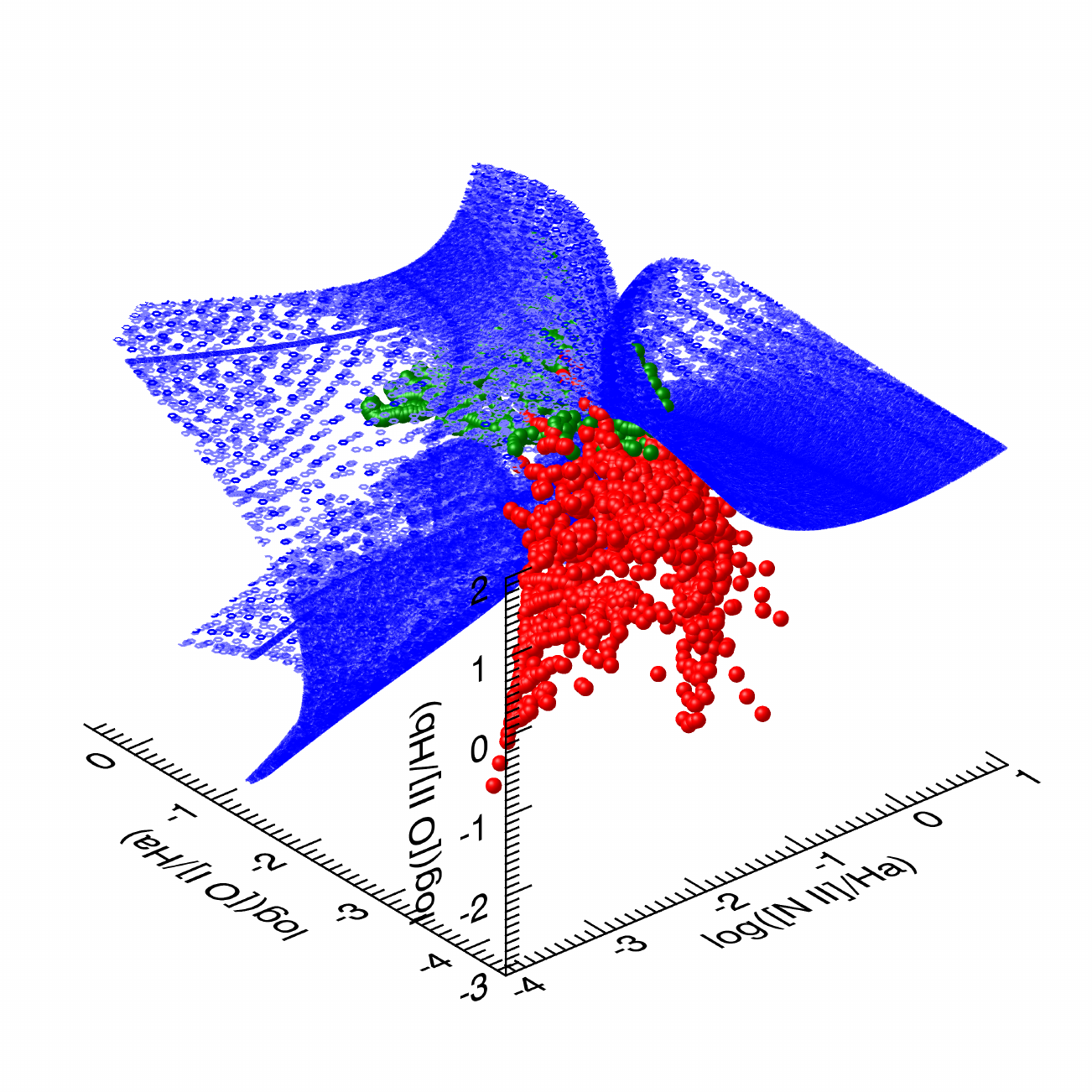}
    \caption{The surface separating shock models \\(SNRs,  green)
from starburst models (HII regions, \\red)  for the  diagnostic $\rm{[N\,II]/H\alpha-[O\,I]/H\alpha-[O\,II]/H\beta}$.}
    \label{fig:NII_OI_OII}
  \end{minipage} 
  \begin{minipage}[b]{0.48\textwidth}
    \includegraphics[width=\textwidth]{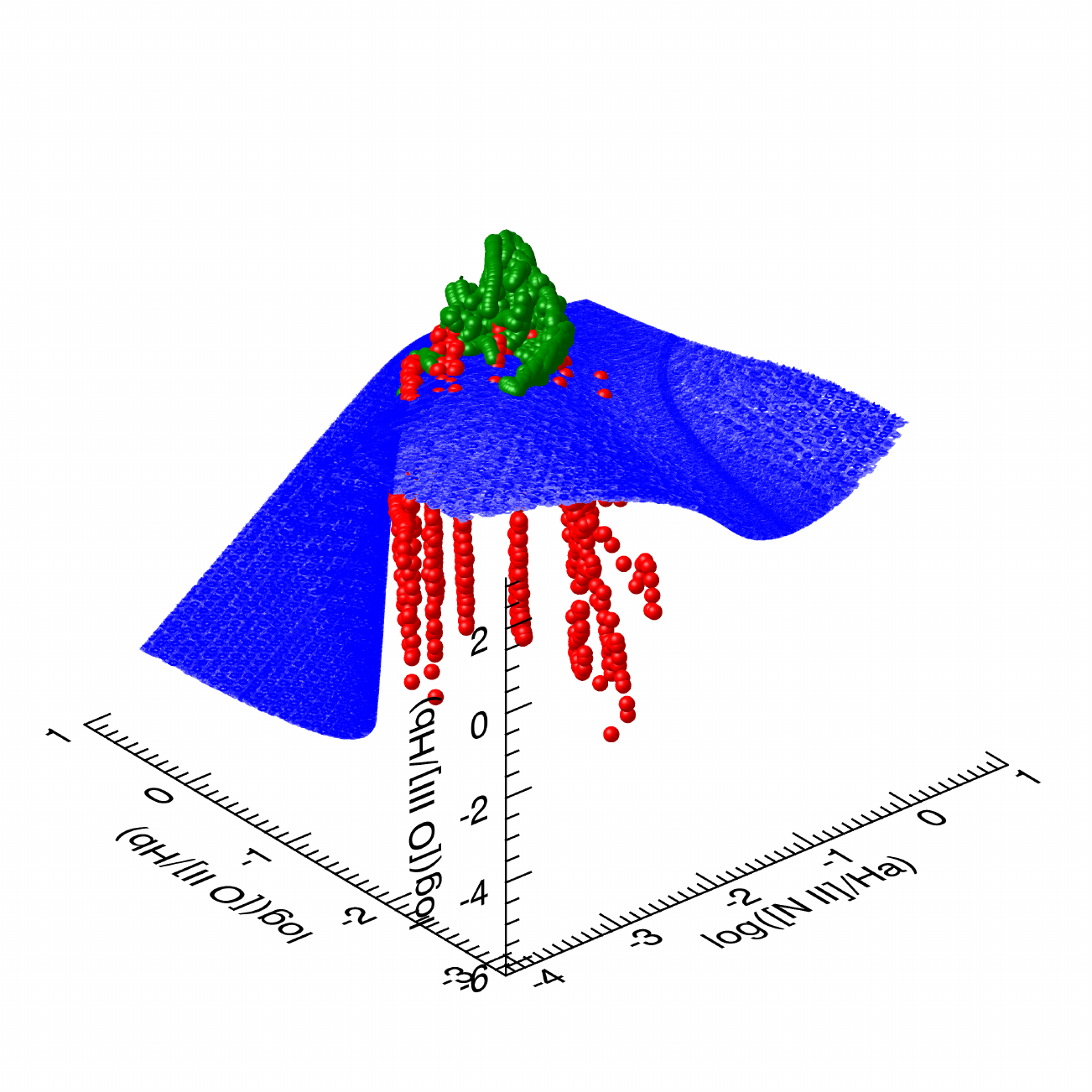}
    \caption{The surface separating shock models \\(SNRs, green)
from starburst models (HII regions, \\red) for the diagnostic $\rm{[N\,II]/H\alpha-[O\,II]/H\beta-[O\,III]/H\beta}$.}
    \label{fig:NII_OII_OIII}
  \end{minipage} 
\end{figure*}

\begin{figure*}
\begin{minipage}[b]{0.48\textwidth}
    \includegraphics[width=\textwidth]{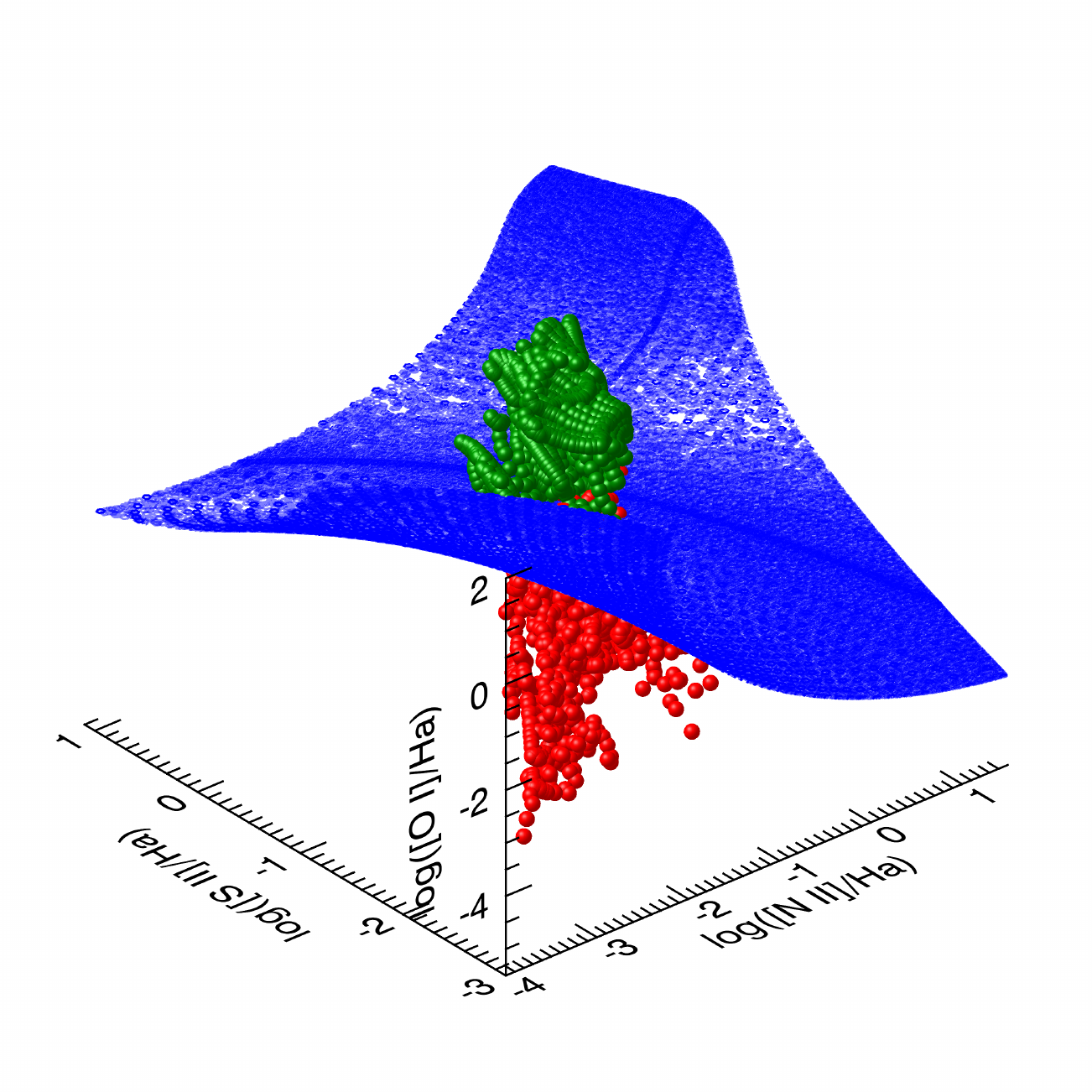}
    \caption{The surface separating shock models \\(SNRs, green)
from starburst models (HII regions,\\ red) for the diagnostic $\rm{[N\,II]/H\alpha-[S\,II]/H\alpha-[O\,I]/H\alpha}$.}
    \label{fig:NII_SII_OI}
  \end{minipage} 
  \begin{minipage}[b]{0.48\textwidth}
    \includegraphics[width=\textwidth]{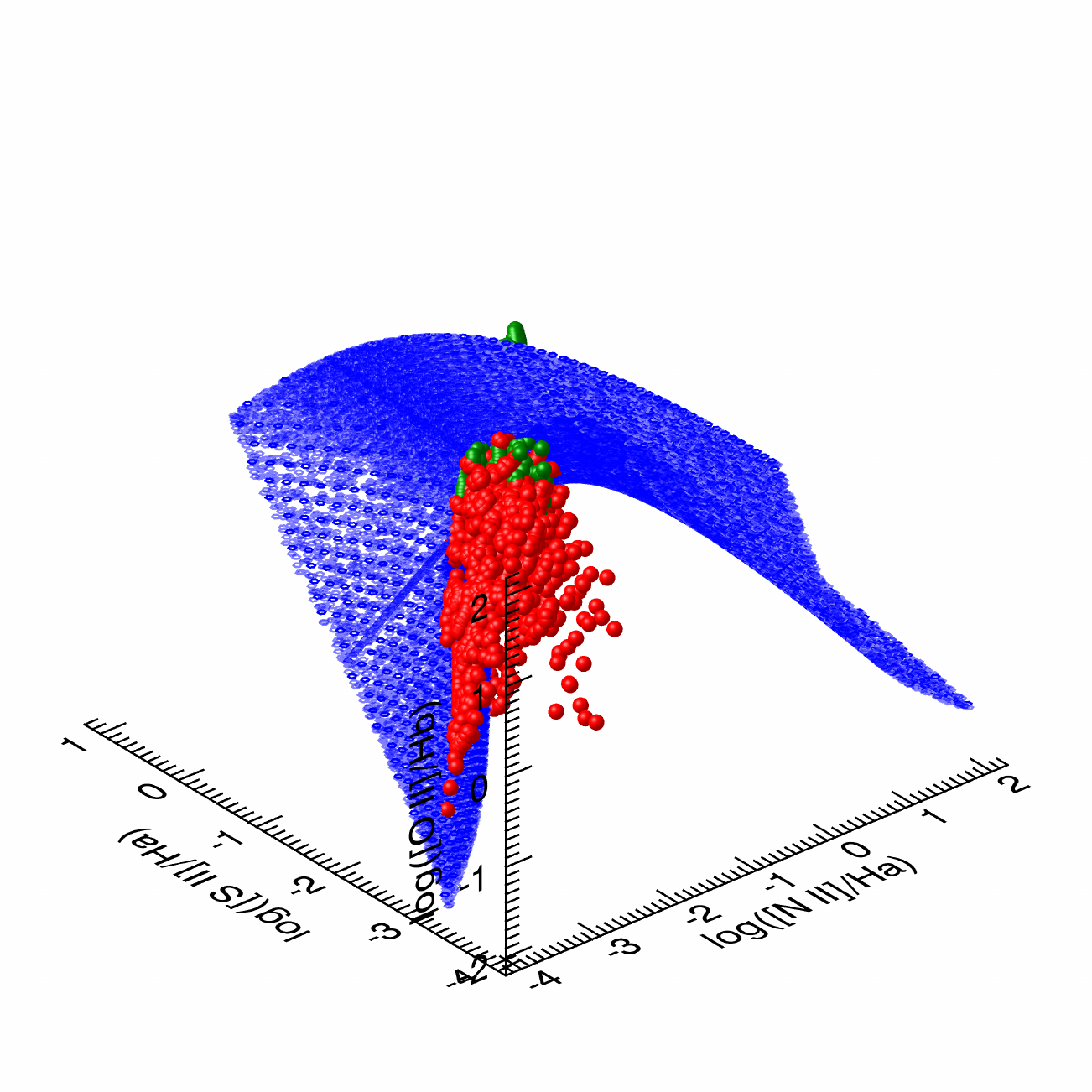}
    \caption{The surface separating shock models \\(SNRs, green)
from starburst models (HII regions, \\red) for the diagnostic$\rm{[N\,II]/H\alpha-[S\,II]/H\alpha-[O\,II]/H\beta}$.}
    \label{fig:NII_SII_OII}
  \end{minipage} 
\end{figure*}

\begin{figure*}
\begin{minipage}[b]{0.48\textwidth}
    \includegraphics[width=\textwidth]{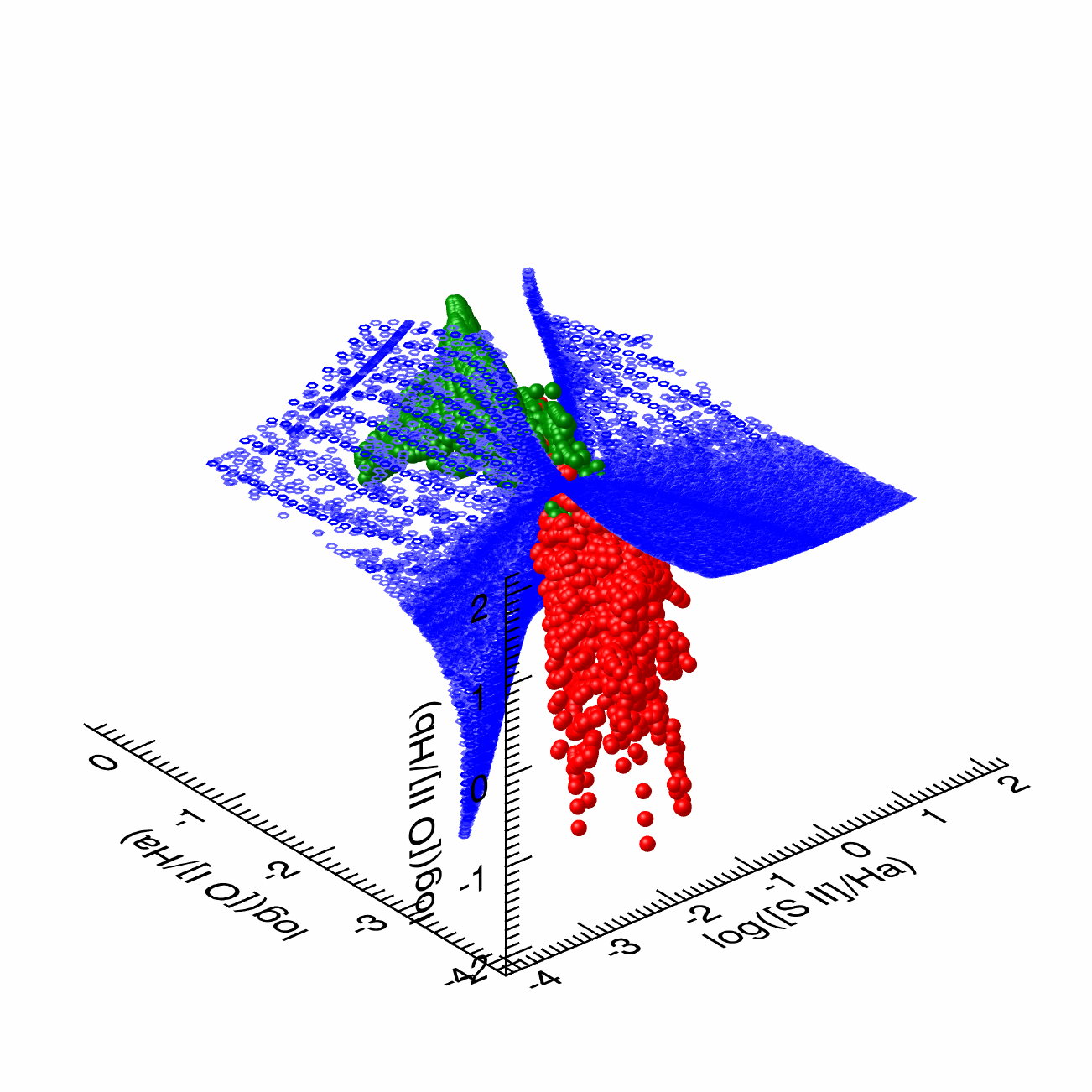}
    \caption{The surface separating shock models \\(SNRs, green)
from starburst models (HII regions, \\red) for the diagnostic $\rm{[S\,II]/H\alpha-[O\,I]/H\alpha-[O\,II]/H\beta}$.}
    \label{fig:SII_OI_OII}
  \end{minipage} 
  \begin{minipage}[b]{0.48\textwidth}
    \includegraphics[width=\textwidth]{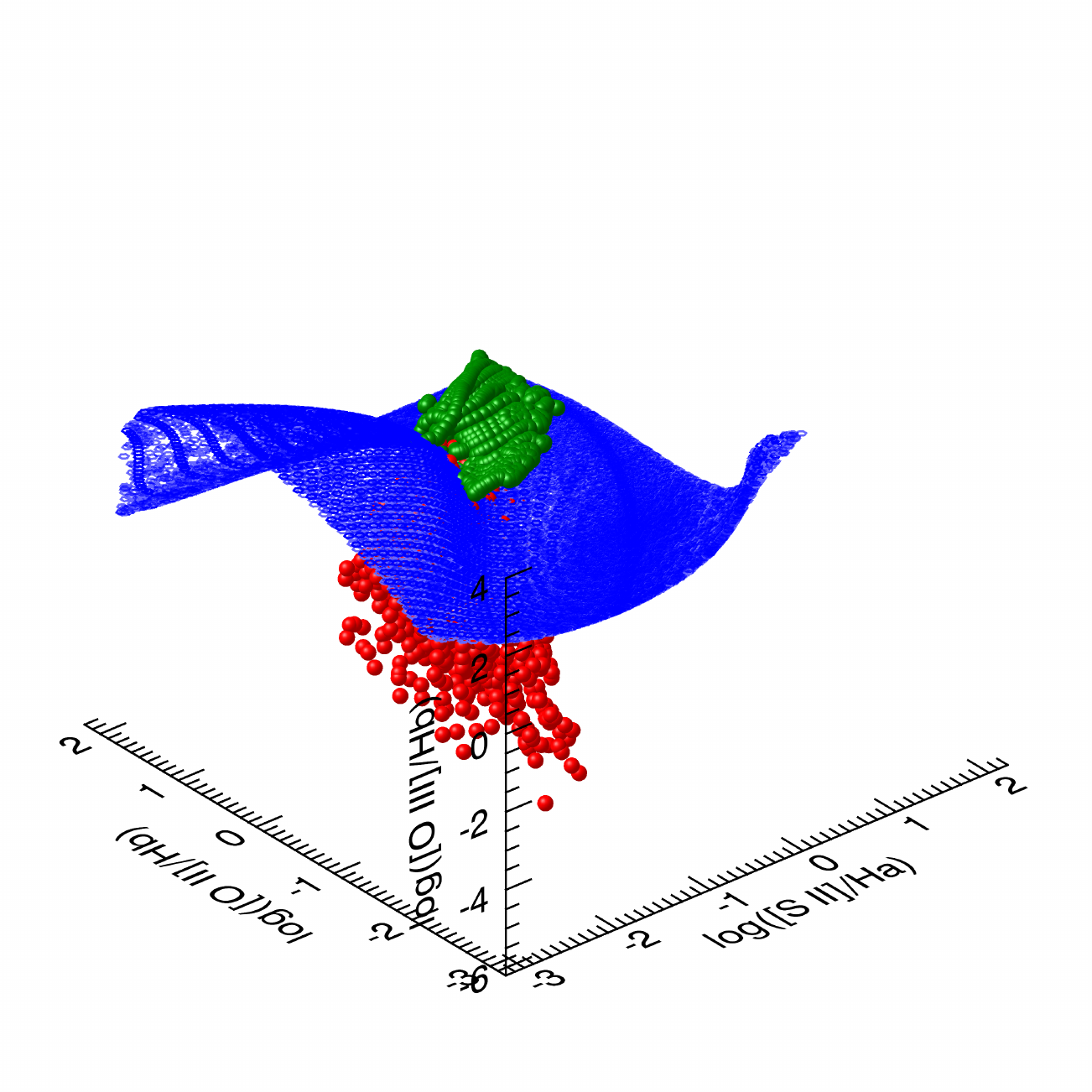}
    \caption{The surface separating shock models \\(SNRs, green)
from starburst models (HII regions, \\red) for the diagnostic $\rm{[S\,II]/H\alpha-[O\,II]/H\beta-[O\,III]/H\beta}$.}
    \label{fig:SII_OII_OIII}
  \end{minipage} 
\end{figure*}

\begin{figure*}
\begin{minipage}[b]{0.48\textwidth}
    \includegraphics[width=\textwidth]{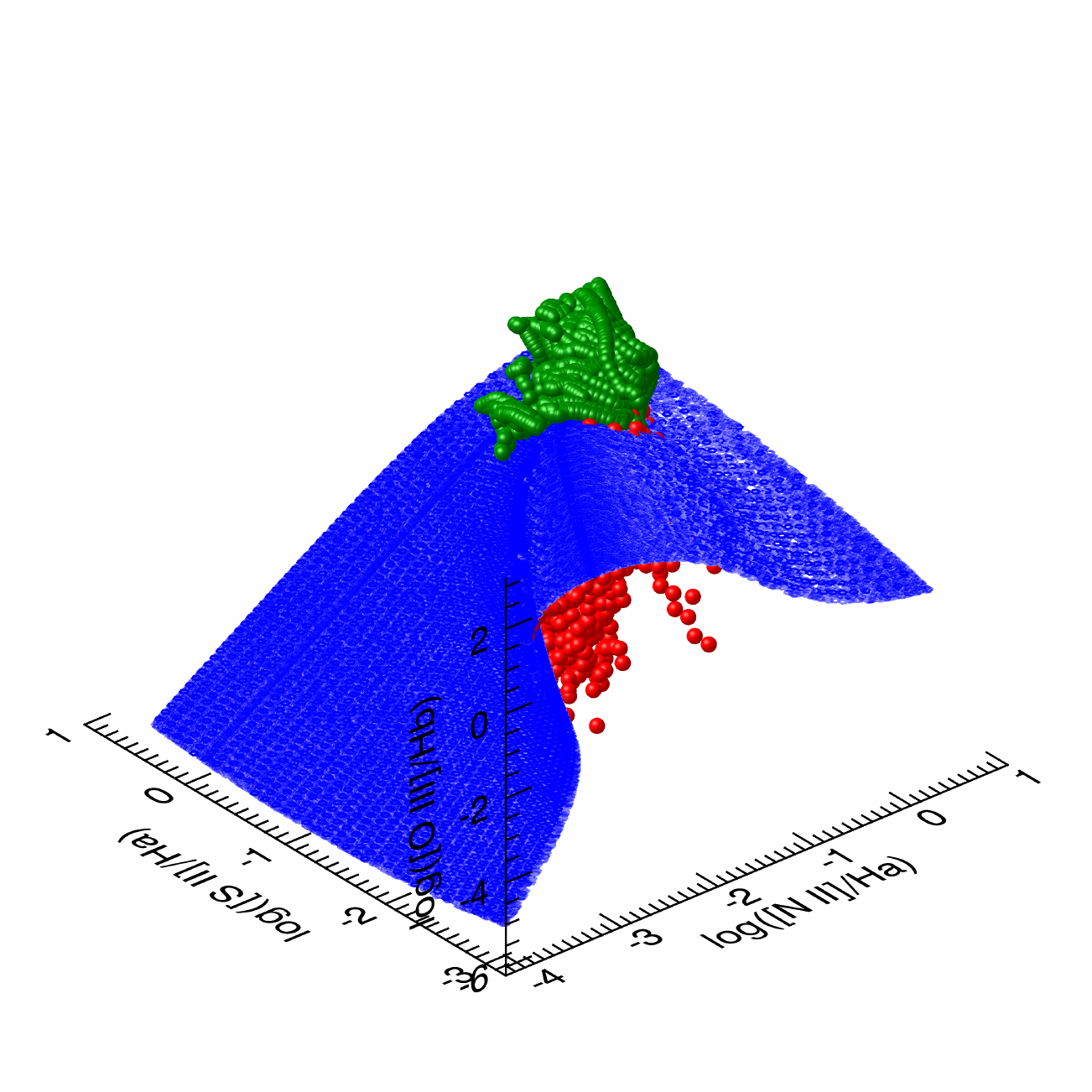}
\caption{The surface separating shock models \\(SNRs, green)
from starburst models (HII regions, \\red) for the diagnostic $\rm{[N\,II]/H\alpha-[S\,II]/H\alpha-[O\,III]/H\beta}$.} 
\label{fig:SII_OII_OIII}
  \end{minipage} 
  \begin{minipage}[b]{0.48\textwidth}
   \includegraphics[width=\textwidth]{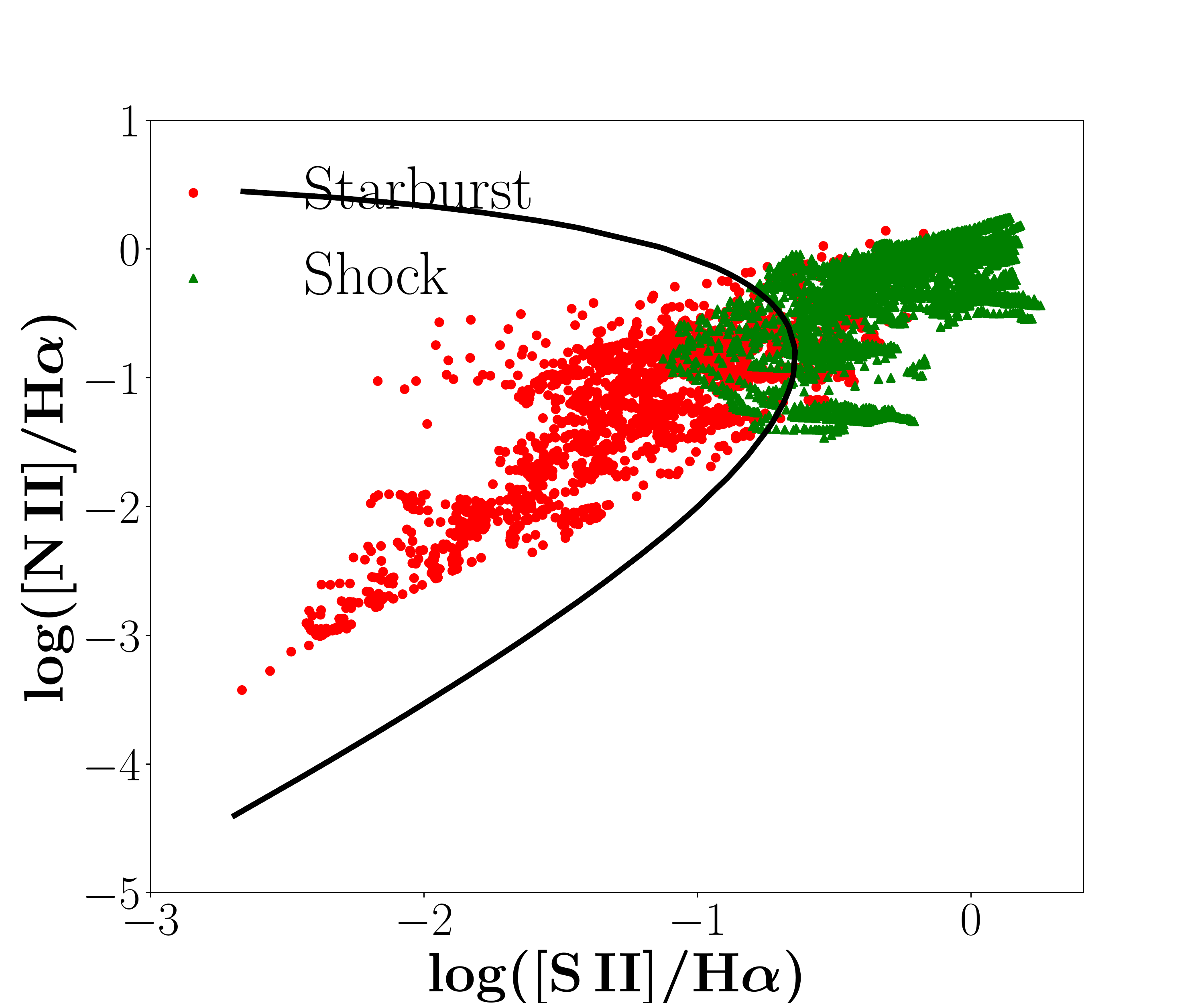}
\caption{The line that separates shock models \\(SNRs, green)
from starburst models (HII regions, \\red) for the diagnostic $\rm{[S\,II]/H\alpha-[N\,II]/H\alpha}$.} 
\label{fig:SII_NII_line}
  \end{minipage} 
\end{figure*}

\begin{figure*}
  \begin{minipage}[b]{0.48\textwidth}
    \includegraphics[width=\textwidth]{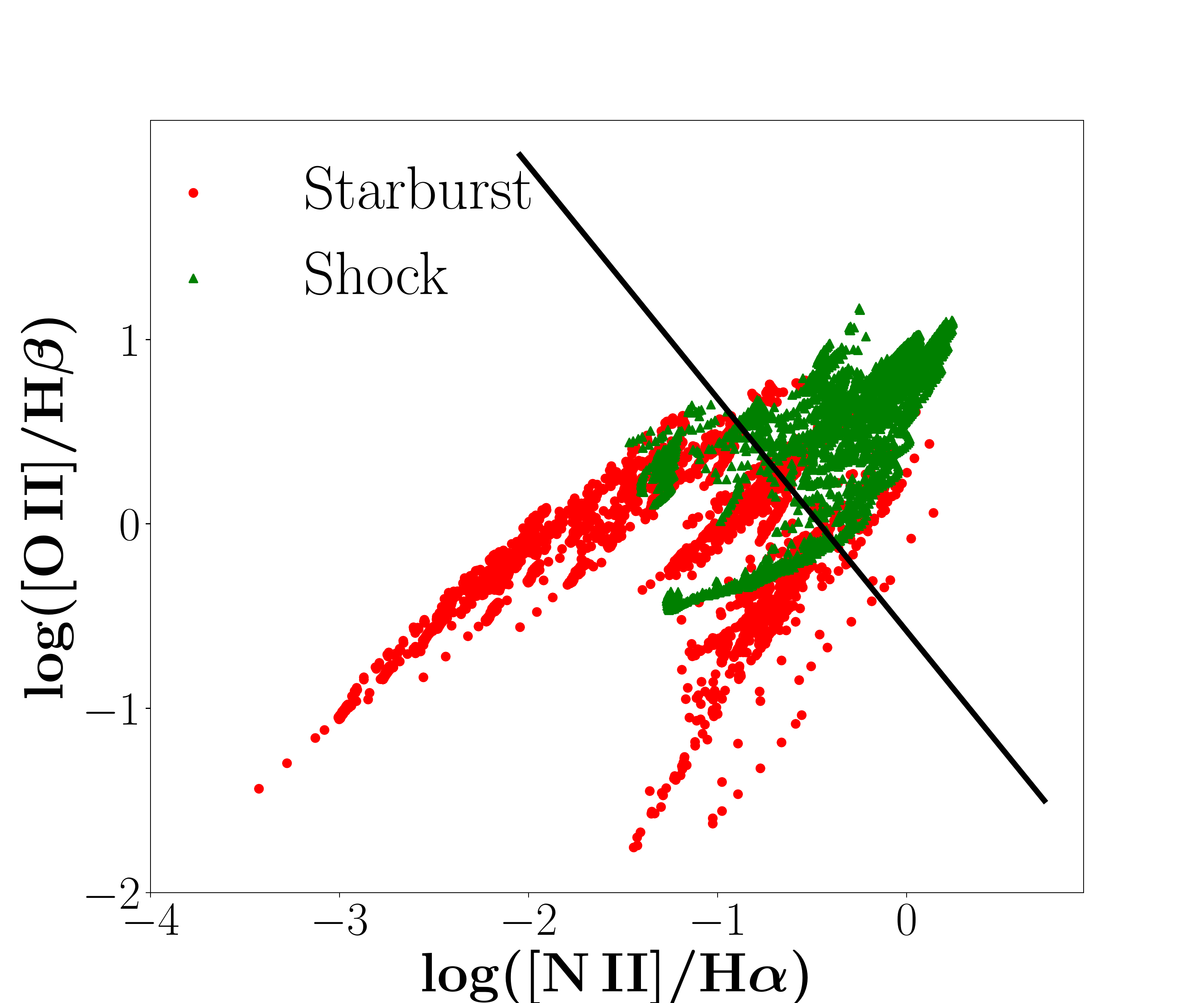}
    \caption{The line that separates shock models \\(SNRs, green)
from starburst models (HII regions, \\red) for the diagnostic $\rm{[N\,II]/H\alpha-[O\,II]/H\beta}$.}
    \label{fig:NII_OII_line}
  \end{minipage}
  \begin{minipage}[b]{0.48\textwidth}
    \includegraphics[width=\textwidth]{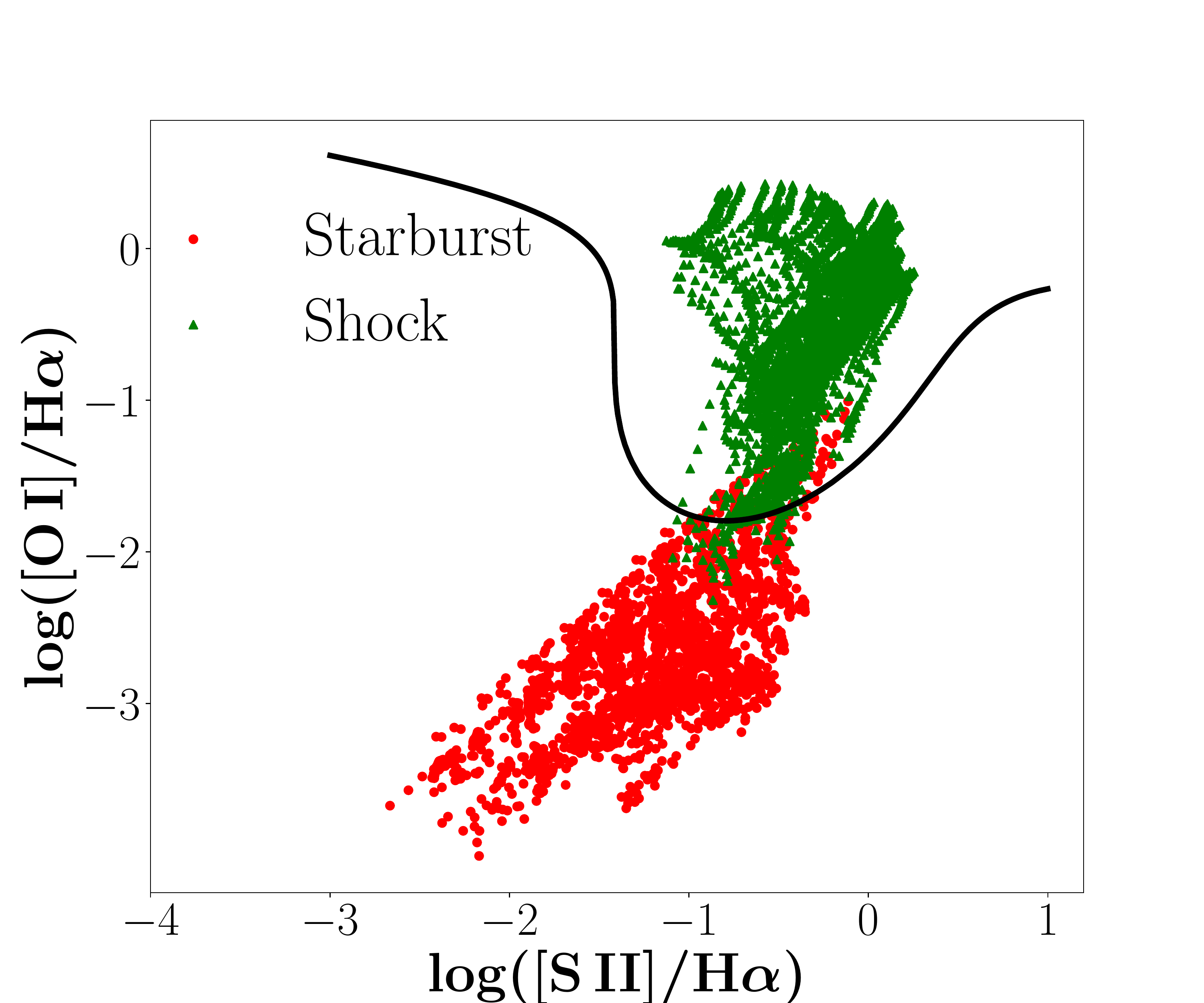}
    \caption{The line that separates shock models \\(SNRs, green)
from starburst models (HII regions, \\red) for the diagnostic $\rm{[S\,II]/H\alpha-[O\,I]/H\alpha}$.}
    \label{fig:SII_OI_line}
  \end{minipage}   
\end{figure*}

\begin{figure*}
  \begin{minipage}[b]{0.48\textwidth}
    \includegraphics[width=\textwidth]{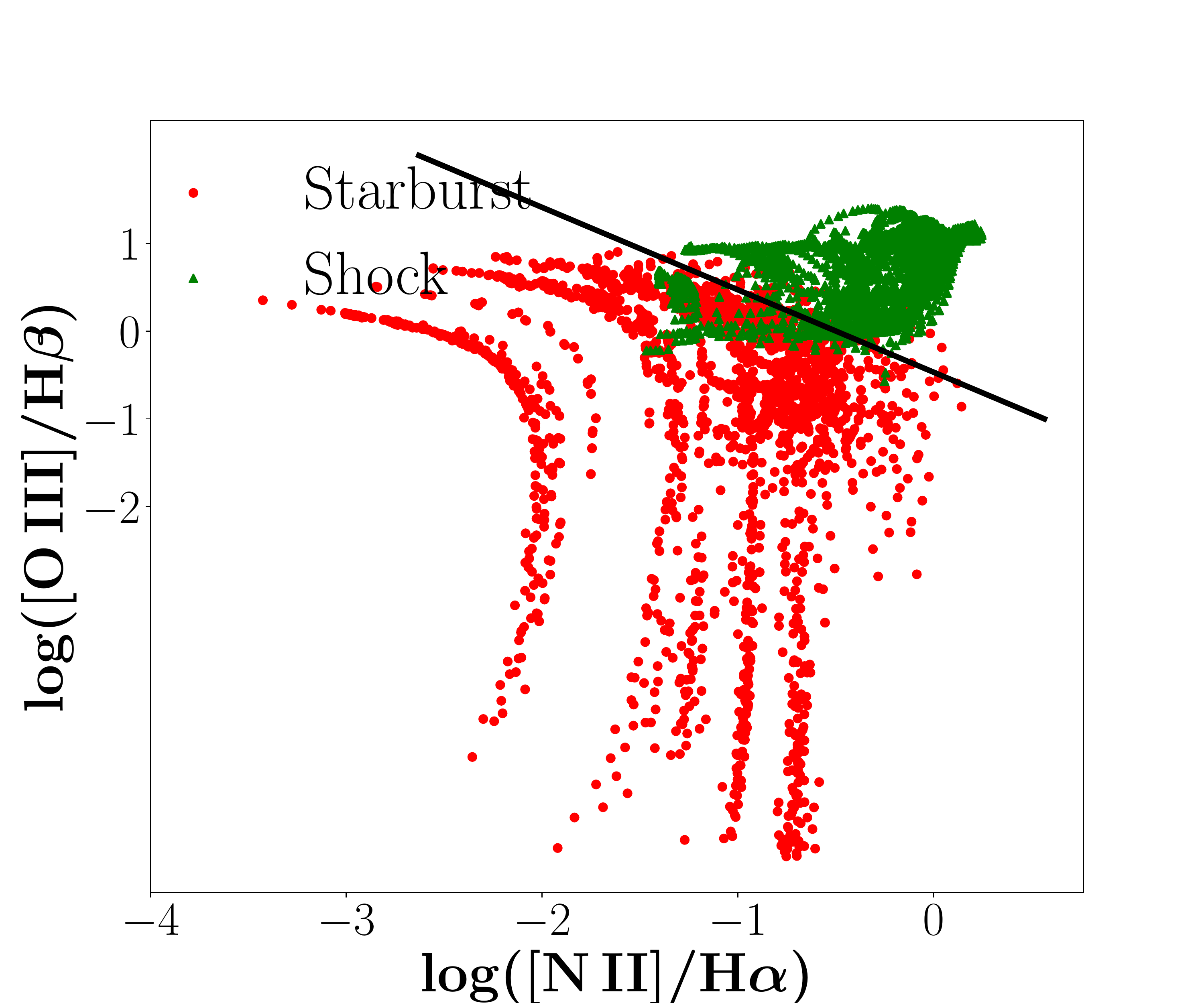}
    \caption{The line that separates shock models \\(SNRs, green)
from starburst models (HII regions, \\red) for the diagnostic $\rm{[N\,II]/H\alpha-[O\,III]/H\beta}$.}
    \label{fig:NII_OIII_line}
  \end{minipage}
  \begin{minipage}[b]{0.48\textwidth}
    \includegraphics[width=\textwidth]{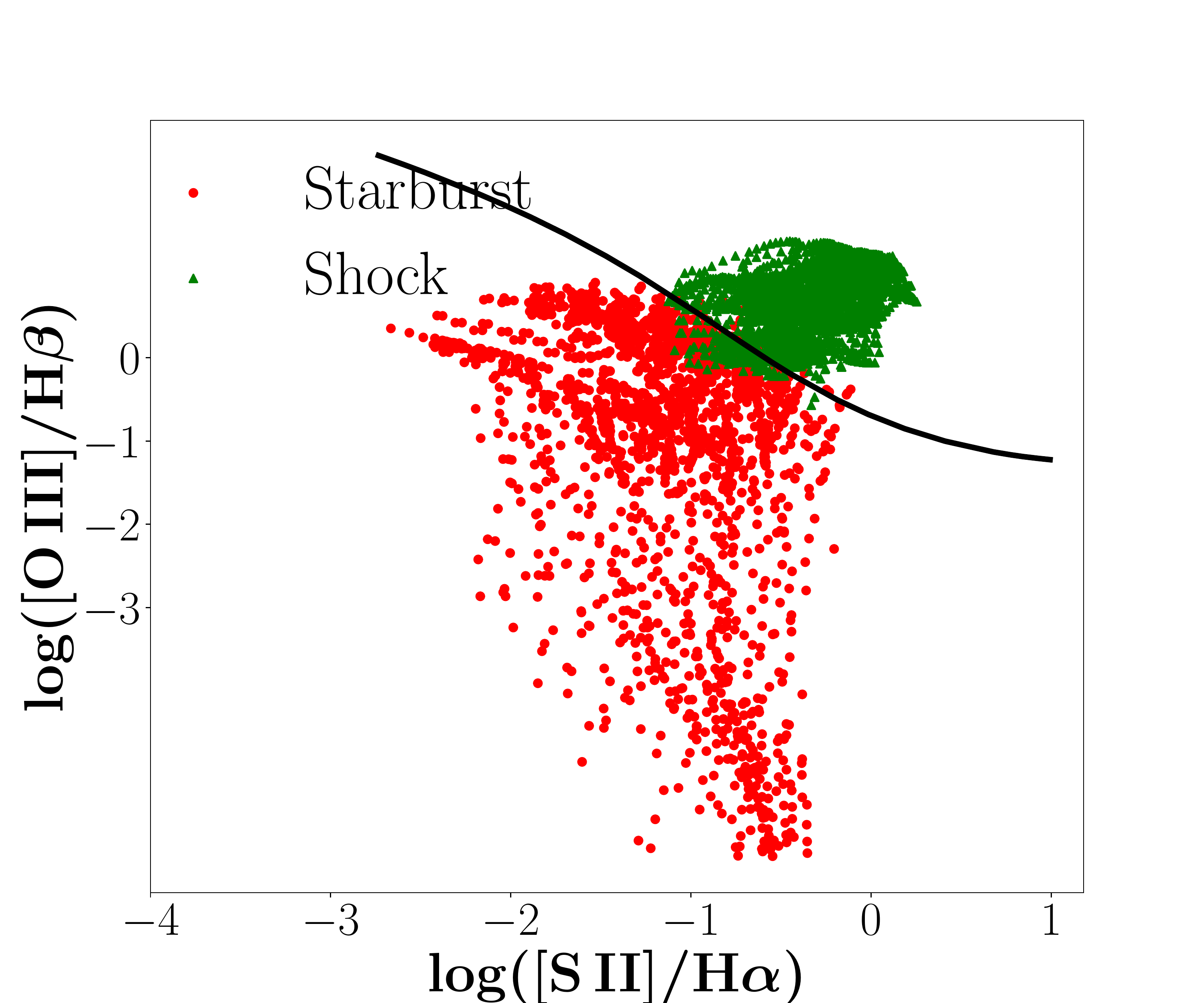}
    \caption{The line that separates shock models \\(SNRs, green)
from starburst models (HII regions, \\red) for the diagnostic $\rm{[S\,II]/H\alpha-[O\,III]/H\beta}$.}
    \label{fig:SII_OIII_line}
  \end{minipage} 
\end{figure*}

\begin{figure*}
  \begin{minipage}[b]{0.48\textwidth}
    \includegraphics[width=\textwidth]{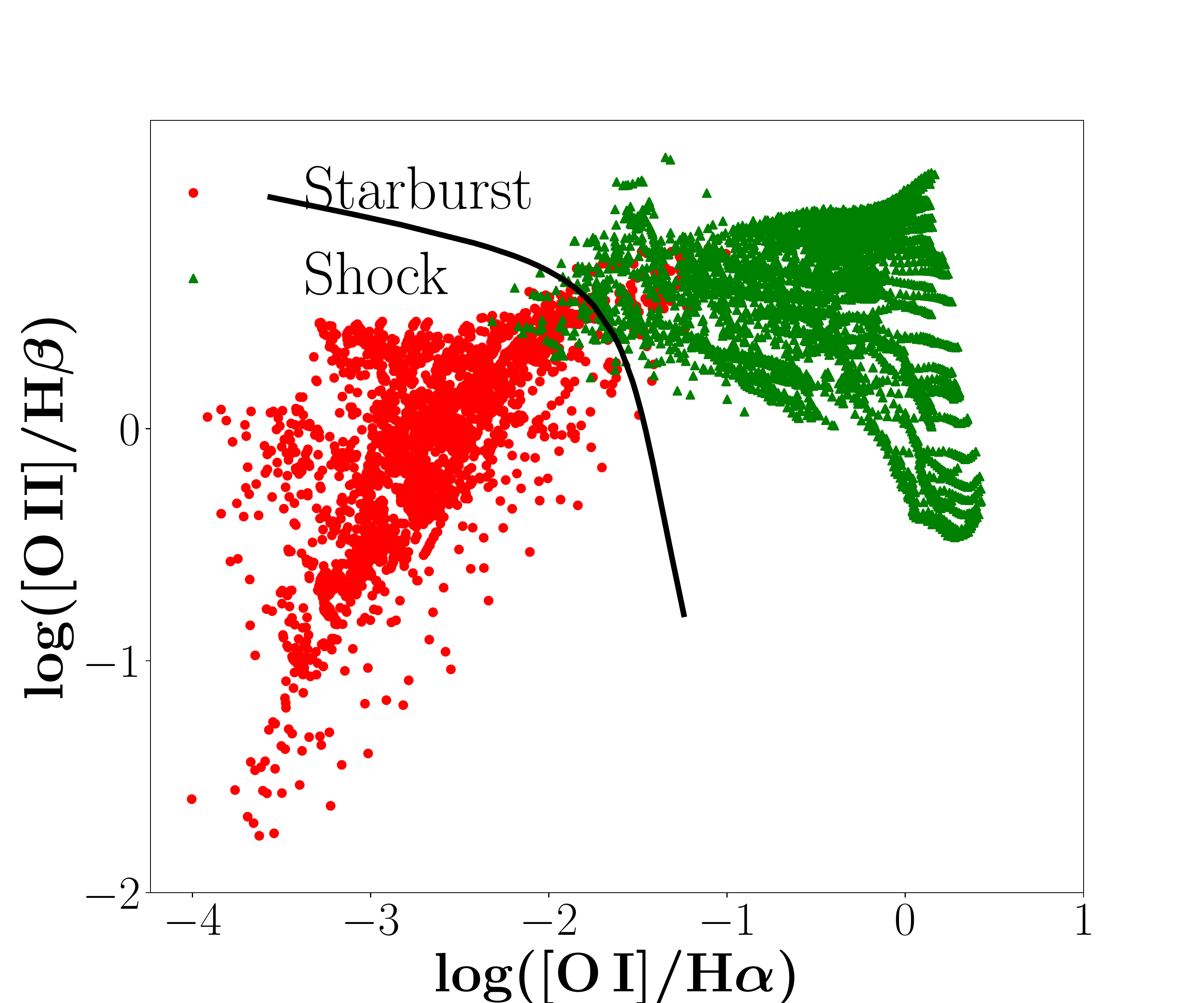}
    \caption{The line that separates shock models \\(SNRs, green)
from starburst models (HII regions, \\red) for the diagnostic $\rm{[O\,I]/H\alpha-[O\,II]/H\beta}$.}
    \label{fig:OI_OII_line}
  \end{minipage}         
  \begin{minipage}[b]{0.48\textwidth}
    \includegraphics[width=\textwidth]{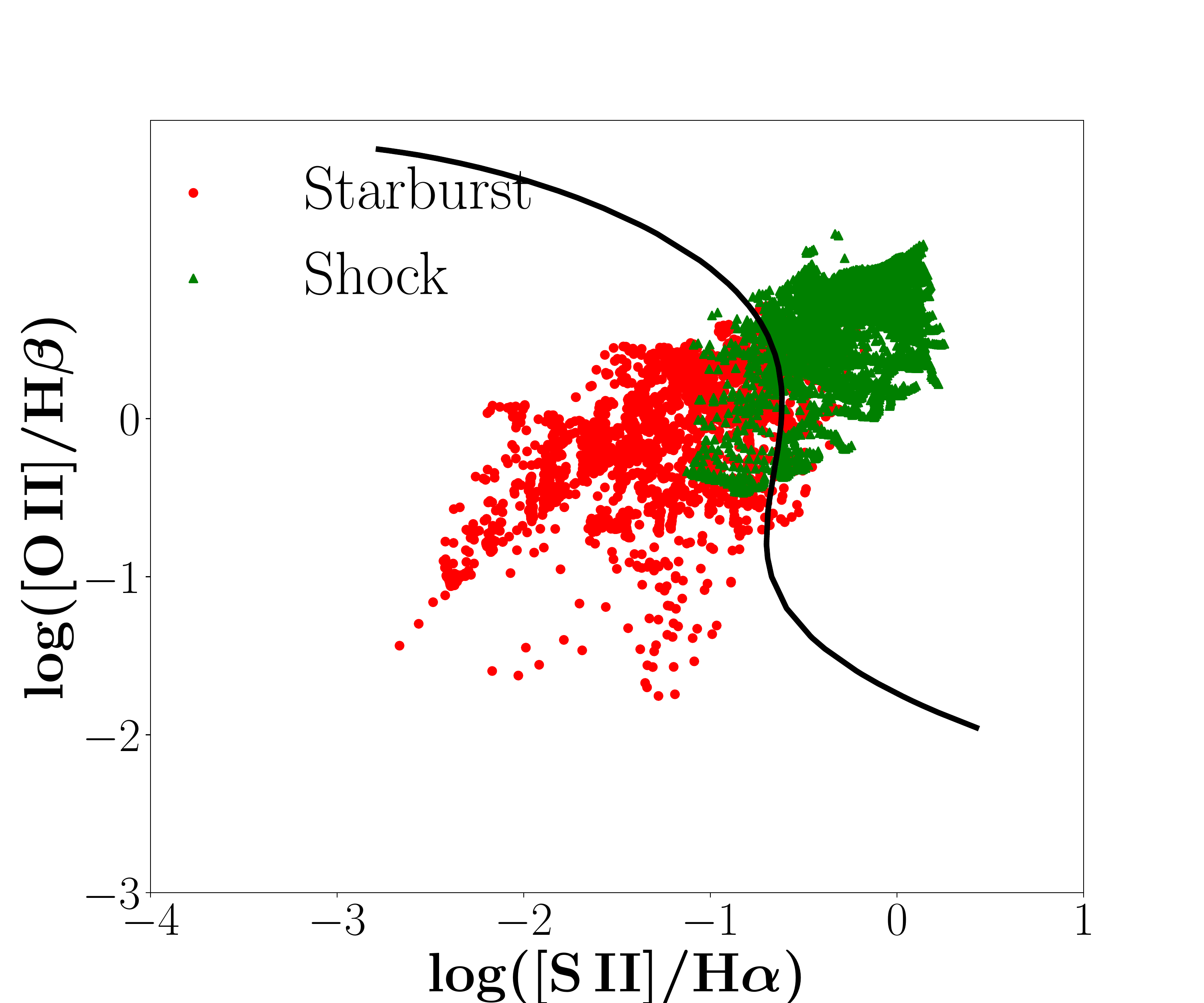}
    \caption{The line that separates shock models \\(SNRs, green)
from starburst models (HII regions, \\red) for the diagnostic $\rm{[S\,II]/H\alpha-[O\,II]/H\beta}$.}
    \label{fig:SII_OII_line}
  \end{minipage}
\end{figure*}

\begin{table*}
\caption{\small{Coefficients of the decision function for the 3D diagnostics.}}
\centering
\begin{threeparttable}
\begin{tabular}{|p{1.5cm}||p{1.8cm}|p{1.8cm}|p{1.8cm}|p{1.8cm}|p{1.8cm}|p{1.8cm}|p{1.8cm}|}
 \hline
 \hfil ijk &\hfil NII-OI-OII &\hfil NII-OII-OIII &\hfil NII-SII-OI &\hfil NII-SII-OII &\hfil NII-SII-OIII &\hfil SII-OI-OII  &\hfil SII-OII-OIII \\
 \hline
 \hfil 000   &\hfil 3.734  &\hfil 3.478   &\hfil 2.382   &\hfil  1.520 &\hfil 2.413 &\hfil 4.108  &\hfil 4.974    \\
 \hfil 010   &\hfil 1.578  &\hfil -1.199  &\hfil -1.822  &\hfil  2.202 &\hfil 1.267 &\hfil 1.213 &\hfil -3.676     \\
 \hfil 020   &\hfil -2.053 &\hfil 2.729   &\hfil 0.151   &\hfil -0.038 &\hfil 0.770 &\hfil -1.274 &\hfil 1.467   \\
 \hfil 030   &\hfil 0.326  &\hfil 0.576   &\hfil -0.002  &\hfil -0.055 &\hfil -1.585 &\hfil 0.826  &\hfil 2.269     \\
 \hfil 001   &\hfil -1.437 &\hfil 5.176   &\hfil 0.214   &\hfil  1.626 &\hfil 4.377 &\hfil -1.758 &\hfil 4.722    \\
 \hfil 011   &\hfil 1.608  &\hfil -1.968  &\hfil 0.658   &\hfil -1.050 &\hfil -2.697 &\hfil 2.587  &\hfil 1.088    \\
 \hfil 021   &\hfil 4.849  &\hfil 1.264   &\hfil -0.021  &\hfil  0.098 &\hfil 1.823 &\hfil 5.000  &\hfil 1.078   \\
 \hfil 002   &\hfil -0.768 &\hfil -2.798  &\hfil -0.804  &\hfil 0.283 &\hfil -2.732 &\hfil -0.747 &\hfil -3.318   \\
 \hfil 012   &\hfil -0.459 &\hfil -2.641   &\hfil 0.014   &\hfil 0.141 &\hfil -1.325 &\hfil -0.094 &\hfil -0.354     \\
 \hfil 003   &\hfil 0.139  &\hfil 3.011   &\hfil 0.223   &\hfil 0.361 &\hfil 1.463  &\hfil 0.266  &\hfil 2.156     \\
 \hfil 100   &\hfil -3.186 &\hfil 7.034   &\hfil -1.821  &\hfil -0.610 &\hfil 0.001 &\hfil -2.305 &\hfil 3.761      \\
 \hfil 110   &\hfil 6.108  &\hfil 3.313   &\hfil -0.795  &\hfil -2.699 &\hfil -6.913 &\hfil 2.439  &\hfil 2.946      \\
 \hfil 120   &\hfil 0.609  &\hfil -0.372  &\hfil  0.127 &\hfil 0.283 &\hfil 1.753 &\hfil -0.096 &\hfil -1.869      \\
 \hfil 101   &\hfil 3.889  &\hfil 0.788   &\hfil  0.149  &\hfil  2.639 &\hfil 1.508 &\hfil 1.342 &\hfil -0.421       \\
 \hfil 111   &\hfil -2.792 &\hfil -0.877  &\hfil  0.379  &\hfil -0.460 &\hfil -2.264 &\hfil -2.236 &\hfil 1.194      \\
 \hfil 102   &\hfil 4.071  &\hfil -3.772  &\hfil  0.196 &\hfil  0.067 &\hfil -0.866 &\hfil 0.762 &\hfil -0.304      \\
 \hfil 200   &\hfil 0.863  &\hfil -1.549  &\hfil -0.350  &\hfil 1.460 &\hfil 0.407 &\hfil -0.172 &\hfil -3.181      \\
 \hfil 210   &\hfil 3.986  &\hfil  8.080  &\hfil  0.250  &\hfil  0.231 &\hfil 1.568 &\hfil 0.359  &\hfil 0.713      \\
 \hfil 201   &\hfil -0.268 &\hfil -0.879  &\hfil  0.253 &\hfil -0.388 &\hfil -0.638 &\hfil -1.707 &\hfil -0.501      \\
 \hfil 300   &\hfil -1.085 &\hfil -1.704  &\hfil  0.025 &\hfil -0.140 &\hfil -2.913 &\hfil 0.042  &\hfil -0.151      \\
\hline
\end{tabular}
\begin{tablenotes}
      \item \scriptsize The line-ratio combinations are presented without the respective Hydrogen lines. In every case:\\
SII = $\rm{[S\,II]/H\alpha}$, NII = $\rm{[N\,II]/H\alpha}$, OI = $\rm{[O\,I]/H\alpha}$,
OII = $\rm{[O\,II]/H\beta}$ and OIII = $\rm{[O\,III]/H\beta}$ . 
\end{tablenotes}
\end{threeparttable}
\label{table:3d_function}
\end{table*}

\begin{table*}
\caption{\small{Coefficients of the decision function for the 2D diagnostics.}}
\centering
\begin{threeparttable}
\begin{tabular}{|p{0.8cm}||p{2cm}|p{2cm}|p{2cm}|p{2cm}|p{2cm}|p{2cm}|p{2cm}|}
 \hline
ij &\hfil NII-OII &\hfil NII-OIII &\hfil OI-OII &\hfil SII-OI &\hfil  SII-OIII &\hfil  SII-OII &\hfil SII-NII \\
 \hline
 00   &\hfil  0.581  &\hfil  0.469   &\hfil 5.017 &\hfil  1.218  &\hfil 3.403   &\hfil 2.777 &\hfil 2.763 \\
 01   &\hfil  1.000   &\hfil  1.000   &\hfil -2.274 &\hfil  0.145  &\hfil  4.433  &\hfil -0.283 &\hfil 1.116 \\
 02   &\hfil 0  &\hfil  0   &\hfil   -1.170  &\hfil 1.371   &\hfil  -0.479  &\hfil 0.981 &\hfil 1.388  \\
 03   &\hfil 0     &\hfil  0   & \hfil  0.432  &\hfil  1.446     &\hfil  0.255   &\hfil 1.185 &\hfil -0.217  \\
 10   &\hfil 1.262     &\hfil 0.939   & \hfil 0.719 &\hfil  -0.449    &\hfil  4.660   &\hfil 4.283 &\hfil 2.244  \\
 11   &\hfil 0     &\hfil  0   &\hfil 3.095 &\hfil  1.932     &\hfil  -0.781   &\hfil -0.332 &\hfil -2.452 \\
 12   &\hfil 0    &\hfil 0   &\hfil 0.124  &\hfil -1.036    &\hfil -0.318   &\hfil -0.225 &\hfil 0.029  \\
 20   &\hfil 0  &\hfil 0   &\hfil -0.842 &\hfil -0.089   &\hfil -1.821  &\hfil -0.575 &\hfil -0.257 \\
 21   &\hfil 0  &\hfil 0   &\hfil 4.291 &\hfil  2.197     &\hfil 0.148  &\hfil -0.511 &\hfil 0.175 \\
 30   &\hfil  0  &\hfil 0   & \hfil 0.701  &\hfil  0.461   &\hfil  0.079   &\hfil -0.403 &\hfil -0.043 \\
\hline
\end{tabular}
\begin{tablenotes}
      \item \scriptsize The line-ratio combinations are presented without the respective Hydrogen lines. In every case:\\
SII = $\rm{[S\,II]/H\alpha}$, NII = $\rm{[N\,II]/H\alpha}$, OI = $\rm{[O\,I]/H\alpha}$,
OII = $\rm{[O\,II]/H\beta}$ and OIII = $\rm{[O\,III]/H\beta}$ .
\end{tablenotes}
\end{threeparttable}
\label{table:2d_function}
\end{table*}

\bsp	
\label{lastpage}
\end{document}